\documentclass[a4paper,11pt]{article}
\pdfoutput=1
\usepackage{jheppub}
\usepackage{amsmath}
\usepackage{hyperref}
\def\bea#1\eea{\begin{align}#1\end{align}}
\def \be  {\begin{equation}}
\def \ee  {\end{equation}}
\newcommand{\nnu}{\nonumber\\}
\newcommand{\kt}{anti-k$_{\rm T}$}
\newcommand{\bef}{\begin{figure}[hbt]\centering}
\newcommand{\eef}{\end{figure}}
\newcommand{\sla}[1]{{#1}\!\!\!\slash}
\newcommand{\R}{{\mathcal R}}
\newcommand\as{\alpha_s}
\newcommand{\f}{\frac}

\title{The semi-inclusive jet function in SCET and small radius resummation for inclusive jet production}

\author{Zhong-Bo Kang,}
\author{Felix Ringer}
\author{and Ivan Vitev}

\affiliation{Theoretical Division, 
                 Los Alamos National Laboratory,
                 Los Alamos, NM 87545, USA}
                                   
\emailAdd{zkang@lanl.gov}
\emailAdd{f.ringer@lanl.gov}
\emailAdd{ivitev@lanl.gov}

\abstract{We introduce a new kind of jet function: the semi-inclusive jet function $J_i(z, \omega_J, \mu)$, 
which describes how a parton $i$ is transformed into a jet with a jet radius $R$ and energy fraction $z = \omega_J/\omega$, with $\omega_J$ and $\omega$ being the large light-cone momentum component of the jet and the corresponding parton $i$ that initiates the jet, respectively. Within the framework of Soft Collinear Effective Theory (SCET) we calculate both $J_q(z, \omega_J, \mu)$ and $J_g(z, \omega_J, \mu)$ to the next-to-leading order (NLO) for cone and \kt~algorithms. We demonstrate that the renormalization group (RG) equations for $J_i(z, \omega_J, \mu)$ follow exactly the usual DGLAP evolution, which can be used to perform the $\ln R$ resummation for {\it inclusive} jet cross sections with a small jet radius $R$. We clarify the difference between our RG equations for $J_i(z, \omega_J, \mu)$ and those for the so-called unmeasured jet functions $J_i(\omega_J, \mu)$, widely used in SCET for {\it exclusive} jet production. Finally, we present applications of the new semi-inclusive jet functions to inclusive jet production in $e^+e^-$ and $pp$ collisions. We demonstrate that single inclusive jet production in these collisions shares the same short-distance hard functions as single inclusive hadron production, with only the fragmentation functions $D_i^h(z, \mu)$ replaced by $J_i(z, \omega_J, \mu)$. This can facilitate more efficient higher-order analytical computations of jet cross sections. We further match our $\ln R$ resummation at both LL$_{R}$ and NLL$_{R}$ to fixed NLO results and present the phenomenological implications for single inclusive jet production at the LHC.}

\begin{document}
\maketitle

\section{Introduction}
\label{sec:intro}
Collimated jets of hadrons play a crucial role in testing the dynamics of the strong interactions and the fundamental properties of Quantum Chromodynamics (QCD)~\cite{Sterman:1977wj,Ellis:2007ib,Buttar:2008jx,Salam:2009jx,Ali:2010tw,Abdesselam:2010pt,Altheimer:2012mn,Sapeta:2015gee,Chien:2015ctp}. They are also one of the main sources for obtaining information about the partonic structure of the nucleon~\cite{Lai:1996mg,Martin:2001es}, for searching for signatures of physics beyond the Standard Model (BSM)~\cite{Stump:2003yu,Butterworth:2008iy}, and for probing the properties of the hot quark gluon plasma created in heavy ion collisions~\cite{Vitev:2008rz,Vitev:2009rd,He:2011pd,Muller:2012zq,Armesto:2015ioy}. Jets are copiously produced at the current highest energy hadron collider, the Large Hadron Collider (LHC) at CERN. Needless to say, reliable predictions of jet cross sections are essential to obtain deeper insights into QCD dynamics, and to constrain any potential signals for BSM physics.

The study of jets requires the use of a jet definition and a jet radius parameter denoted by $R$~\cite{Cacciari:2011ma,Salam:2007xv}, which determines how close in angle two particles have to be in order to be clustered into the same jet. Many jet and jet substructure observables have been resummed to very high accuracy within the powerful framework of Soft Collinear Effective Theory (SCET)~\cite{Bauer:2000ew,Bauer:2000yr,Bauer:2001ct,Bauer:2001yt}. One class of logarithms, to be resummed for jet production, that is under active discussion at the moment are logarithms of the jet radius parameter, $\ln R$. When the jet radius $R$ is small, such  logarithms can become large, thus potentially impacting the convergence of the conventional perturbative expansion in terms of the strong coupling constant $\alpha_s$  and requiring resummation. Such resummation is highly desirable, since there is a growing  use of small $R$ values in jet observables and/or modern jet analysis, especially for jet substructure. Smaller jet radii, as small as $R=0.2$, are also commonly used in heavy ion collisions~\cite{Aad:2012vca,Abelev:2013kqa,Aad:2014wha,Chatrchyan:2014ava,Adam:2015ewa} in order to reduce the effects of fluctuations in the heavy-ion background.

For narrow jets, resummation of logarithms of the jet radius $R$ for the jet cross sections is one of the hot topics discussed actively in the QCD community at the moment. The $\ln R$ resummation has been studied by several groups within SCET, see, e.g., Refs.~\cite{Becher:2015hka,Chien:2015cka,Becher:2016mmh,Kolodrubetz:2016dzb}, where generally Sudakov double logarithms of the jet radius arise. In particular, the associated jet function in these processes has a 
$\left(\alpha_s \ln^2 R\right)^n$ dependence. On the other hand, Dasgupta, Dreyer, Salam and Soyez also discussed the resummation of  the jet radius parameter at leading logarithmic order in~\cite{Dasgupta:2014yra,Dasgupta:2016bnd}, which exhibits single logarithms of the form $\left(\alpha_s \ln R\right)^n$. At the same time, the explicit next-to-leading order (NLO) calculations for single inclusive jet cross section exhibit a single logarithmic dependence on $R$~\cite{Jager:2004jh,Mukherjee:2012uz}. Such an apparently different structure of the logarithmic dependence on the jet radius $R$ has been noticed before~\cite{Dasgupta:2014yra,Kolodrubetz:2016dzb}. We further illuminate this important issue from a different perspective. 

In this paper, within the framework of SCET, we introduce a new jet function -- the semi-inclusive jet function $J_i(z, \omega_J, \mu)$, which describes a jet with energy $\omega_J$ and radius $R$, carrying a fraction $z$ of the large light-cone momentum component of the parton $i$ that initiates the jet~\cite{KRV}. We demonstrate that  these semi-inclusive jet functions  are the ones relevant to the calculations of {\it inclusive} jet cross sections. We calculate $J_i(z, \omega_J, \mu)$ for both quark and gluon jets to NLO accuracy. We demonstrate that the renormalization group (RG) equations for $J_i(z, \omega_J, \mu)$ follow exactly the usual timelike DGLAP evolution~\cite{Gribov:1972ri,Lipatov:1974qm,Dokshitzer:1977sg,Altarelli:1977zs}, which can be used to perform the $\ln R$ resummation for {\it inclusive} jet cross sections with a small jet radius $R$. We clarify the difference between our RG equations for $J_i(z, \omega_J, \mu)$ and those for the  so-called unmeasured jet functions $J_i(\omega_J, \mu)$, widely used in SCET for {\it exclusive} jet productions. In other words, the aforementioned single and double logarithm differences are simply due to the difference in the jet observables, {\it inclusive} vs {\it exclusive} jet cross sections.

In addition, we present  applications of the semi-inclusive jet functions to  single inclusive jet production in $e^+e^-$ and $pp$ collisions: $e^+  e^-\to {\rm jet} X$ and $p p\to {\rm jet} X$. We demonstrate that single inclusive jet production in these collisions shares the same short-distance hard functions as single inclusive hadron production, $e^+e^-\to h X$ and $p p\to h X$, with only the fragmentation functions $D_i^h(z, \mu)$ replaced by $J_i(z, \omega_J, \mu)$. We expect that this finding will facilitate more efficient higher-order computations of jet cross sections~\cite{deFlorian:2007fv,Ridder:2013mf,deFlorian:2013qia}, as one can evaluate the individual pieces separately. The semi-inclusive jet functions can also be used in the study of jet physics in $e p$ collisions at an electron ion collider (EIC)~\cite{Kang:2012zr,Kang:2013nha,Kang:2013wca,Kang:2013lga,Kang:2014qba,Hinderer:2015hra,Boer:2011fh,Accardi:2012qut}.

The rest of the paper is organized as follows. In Sec.~\ref{sec:definition} we  set up the theoretical framework and give the SCET definitions of the semi-inclusive jet functions. We compute the NLO semi-inclusive jet functions for both quark and gluons jets, and derive their renormalization group equations. At the end of this section, we also present the numerical solution of the RG equations and obtain the evolved semi-inclusive jet functions. In Sec.~\ref{sec:ee}, using $e^+e^-$ collisions as an example, we present the factorized cross sections for $e^+  e^-\to h X$ and $e^+  e^-\to {\rm jet} X$. We compute the NLO hard functions, and demonstrate that they are the same for single inclusive hadron/jet production. In Sec.~\ref{sec:pp} we generalize the factorized formalism in $e^+e^-$ collisions to $pp$ collisions and present in detail the phenomenological implications of the $\ln R$ resummation for single inclusive jet production at the LHC. We conclude our paper in Sec.~\ref{sec:summary}.

\section{The semi-inclusive jet function}
\label{sec:definition}
In this section we start by setting up the theoretical framework for our analysis and introduce the relevant SCET ingredients. We then give the definition of the semi-inclusive quark and gluon jet functions in SCET and calculate  them to next-to-leading order. From the explicit calculations, we discuss their renormalization and how the corresponding renormalization group equations can be used to achieve small jet radius resummation for inclusive jet spectra. 

\subsection{SCET ingredients}
SCET~\cite{Bauer:2000ew,Bauer:2000yr,Bauer:2001ct,Bauer:2001yt} is an effective  theory of QCD, describing the interactions of soft and collinear degrees of freedom in the presence of hard scattering.  It has been successfully applied to study  a wide variety of hard scattering processes at the LHC, especially jet producion. Jets are collimated spray of hadrons, and are conveniently described using light-cone coordinates. Typically, we introduce a light-cone vector $n^\mu$ whose spatial part is along the jet axis, and another conjugate vector $\bar n^\mu$ such that $n^2 = \bar n^2 = 0$ and $n\cdot \bar n = 2$. Any four-vector $p^\mu$ can then be decomposed as $p^\mu = (p^+, p^-, p_\perp)$ with $p^+ = n\cdot p$, $p^- = \bar n\cdot p$. In other words,
\bea
p^\mu = p^- \frac{n^\mu}{2} + p^+ \frac{\bar n^\mu}{2} + p_\perp^\mu. 
\eea
The momentum $p^\mu$ of a particle within a jet scales collinearly, with $p^\mu= (p^+, p^-, p_\perp) \sim p^- (\lambda^2, 1, \lambda)$. 

The gauge invariant quark and gluon fields are given by
\bea
\chi_{n} =  W_n^\dagger \xi_n,
\qquad
{\mathcal B}_{n\perp}^\mu =  \frac{1}{g}\left[W_n^\dagger iD_{n\perp}^\mu W_n\right],
\eea 
and are composite SCET fields of $n$-collinear quarks and gluons. Here $iD_{n\perp}^\mu = {\mathcal P}_{n\perp}^\mu + gA_{n\perp}^\mu$, and ${\mathcal P}^\mu$ is the label momentum operator. On the other hand, $W_n$ is the Wilson line of collinear gluons,
\bea
W_n(x) = \sum_{\rm perms} \exp\left[-g\frac{1}{\bar n\cdot {\mathcal P}} \bar n\cdot A_n(x)\right].
\eea
We further define
\bea
\chi_{n,\omega} = \delta\left(\omega - \bar n\cdot {\mathcal P}\right) \chi_n,
\qquad
{\mathcal B}_{n\perp, \omega}^\mu = \delta\left(\omega - \bar n\cdot {\mathcal P}\right) {\mathcal B}_{n\perp}^\mu . 
\eea
At leading order in the SCET power expansion, the interactions of soft gluons with collinear quark/gluon fields exponentiate to form eikonal Wilson lines. One might redefine the above collinear fields to decouple collinear-soft interactions in the Lagrangian~\cite{Bauer:2001yt}. In the rest of the paper, all the collinear fields $\chi_{n}$ and ${\mathcal B}_{n\perp}^\mu$ are understood to be those after the field redefinition, and thus do not interact with soft gluons.

\subsection{Definition and jet algorithms}
With the above gauge invariant quark and gluon fields, we can construct the following semi-inclusive quark and gluon jet functions $J_q(z, \omega_J)$ and $J_g(z, \omega_J)$, respectively
\bea
J_q(z = \omega_J / \omega, \omega_J, \mu) &= \frac{z}{2N_c}{\rm Tr} \left[\frac{\sla{\bar n}}{2}
\langle 0| \delta\left(\omega - \bar n\cdot {\mathcal P} \right) \chi_n(0)  |JX\rangle \langle JX|\bar \chi_n(0) |0\rangle \right],
\\
J_g(z = \omega_J / \omega, \omega_J, \mu) &= - \frac{z\,\omega}{2(N_c^2-1)}
\langle 0|  \delta\left(\omega - \bar n\cdot {\mathcal P} \right) {\mathcal B}_{n\perp \mu}(0) 
 |JX\rangle \langle JX|{\mathcal B}_{n\perp}^\mu(0)  |0\rangle,
\eea
where the state $|JX\rangle$ represents the final-state unobserved particles $X$ and the observed jet $J$. Note that summation over the unobserved particles $X$ is implied, and $\omega_J= \bar n\cdot p_J$ is the large light-cone momentum component of the jet with  momentum $p_J$. On the other hand, $\omega$ is the large light-cone momentum component of the parton (either $q$ or $g$) which initiates the jet. We will refer to $\omega_J$ and  $\omega$ as energy for simplicity in the rest of the paper. Our semi-inclusive jet functions $J_i(z, \omega_J)$ can thus be interpreted as the probability of the parton $i$ with energy $\omega$ to transform into a jet with energy $\omega_J =z\, \omega$. In some sense, this is similar to the so-called microjet fragmentation function introduced in \cite{Dasgupta:2014yra,Dasgupta:2016bnd}. They are very similar to the usual quark and gluon fragmentation functions, which are defined as follows
\bea
D_q^h(z = p_h^{-} / \omega, \mu) &= \frac{z}{2N_c}{\rm Tr} \left[\frac{\sla{\bar n}}{2}
\langle 0|  \delta\left(\omega - \bar n\cdot {\mathcal P} \right) \chi_n(0) |hX\rangle \langle hX|\bar \chi_n(0) |0\rangle \right],
\\
D_g^h(z = p_h^{-} / \omega, \mu) &= - \frac{z\,\omega}{2(N_c^2-1)}
\langle 0|  \delta\left(\omega - \bar n\cdot {\mathcal P} \right) {\mathcal B}_{n\perp \mu}(0) 
 |hX\rangle \langle hX|{\mathcal B}_{n\perp}^\mu(0)  |0\rangle.
\eea

We will now calculate the semi-inclusive jet function for both quark and gluon initiated jets. We start with $J_q(z, \omega_J)$, where we present detailed derivations. For $J_g(z, \omega_J)$, the calculation is similar, and we present only the final results. At leading order (LO), the results are simple, we have
\bea
J_q^{(0)}(z, \omega_J) &= \delta(1-z),
\\
J_g^{(0)}(z, \omega_J) &= \delta(1-z),
\eea
where the superscript $(0)$ represents the LO result.

At next-to-leading order, the results of jet functions depend on the jet algorithm. For example, at the LHC a longitudinally-invariant $k_T$-type algorithm is usually used~\cite{Cacciari:2011ma}, which introduces a distance between every pair of particles $i$ and $j$
\bea
d_{ij} = {\rm min} \left(p_{T i}^{2p},\, p_{T j}^{2p}\right) \frac{\Delta R_{ij}^2}{R^2},
\eea
and a distance measure between each particle and the beam,
\bea
d_{iB} = p_{Ti}^{2p}.
\eea
Here $p=1,0,-1$ correspond to the ${\rm k_T}$,  Cambridge/Aachen, and \kt~algorithm, respectively. $R$ is the jet radius parameter, and
\bea
\Delta R_{ij} = \sqrt{\left(\Delta \eta_{ij}\right)^2 + \left(\Delta \phi_{ij}\right)^2},
\eea
where $\Delta \eta_{ij}$ and $\Delta \phi_{ij}$ are the rapidity and azimuthal differences between the particles $i$ and $j$. The algorithm proceeds by identifying the smallest of the $d_{ij}$ and $d_{iB}$. If it is a beam distance $d_{iB}$, the particle $i$ is defined as a jet and removed from the list of particles. If the smallest distance is a $d_{ij}$, the two particles $i$, $j$ are merged into a single one. The procedure is repeated until no particles are left in the event. In the so-called narrow jet approximation~\cite{Jager:2004jh,Mukherjee:2012uz}, where all the particles in the jet are collimated along the jet axis, one can show~\cite{Hornig:2016ahz} that the jet algorithm constraint amounts to 
\bea
\Delta R_{ij} \approx \theta_{ij} \cosh \eta < R,
\eea
or equivalently 
\bea
\theta_{ij} < \frac{R}{\cosh \eta} \equiv \mathcal{R},
\label{eq:algorithm}
\eea
where $\theta_{ij}$ is the angle between particles $i$ and $j$, and $\eta$ is the jet rapidity. On the other hand, for the cone-jet algorithm~\cite{Salam:2007xv} the constraint will be different and it leads to 
\bea
\theta_{iJ} < \R,
\eea
where $\theta_{iJ}$ is the angle between the jet and the particle $i$ that belongs to the jet. For detailed discussion, see Ref.~\cite{Ellis:2010rwa}. 

\bef
\includegraphics[width=\textwidth]{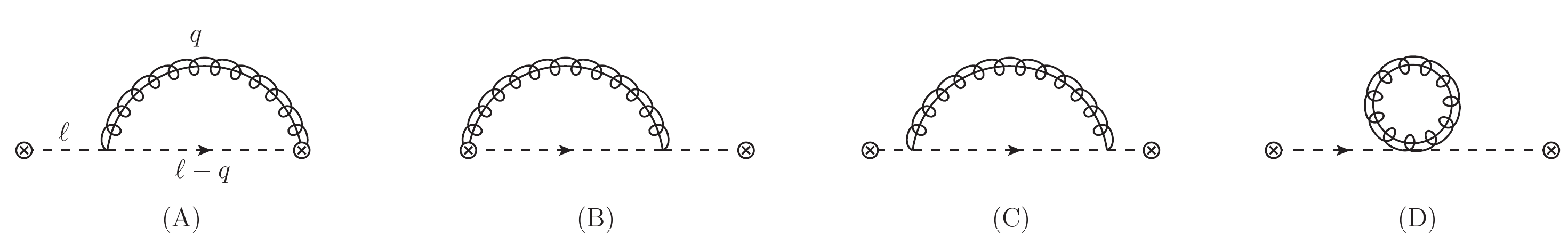}
\caption{Feynman diagrams that contribute to the semi-inclusive quark jet function. The quark that initiates the jet has momentum $\ell = (\ell^- = \omega, \ell^+, 0_\perp)$, with $\omega = \omega_J/z$ and $\omega_J$ the jet energy.}
\label{fig:Jq}
\eef

\subsection{The semi-inclusive quark jet function}
Let us now turn to the detailed calculations for the semi-inclusive quark jet function. The Feynman diagrams which contribute to $J_q(z, \omega_J)$ are given in Fig.~\ref{fig:Jq}, where an incoming quark with momentum $\ell = (\ell^- = \omega, \ell^+, 0_\perp)$ splits into a gluon $q=(q^-, q^+, q_\perp)$ and a quark $\ell-q = (\omega - q^-, \ell^+ - q^+, -q_\perp)$. The total forward scattering matrix element can be computed as a sum over all cuts. The only diagrams that contribute are the cuts through the loops, where there are two final-state partons. All the virtual diagrams which correspond to the cuts through only one parton lead to scaleless integrals and thus vanish in dimensional regularization (via $1/\epsilon_{\rm UV} - 1/\epsilon_{\rm IR} = 0$). It would be possible to separate IR and UV singularities by e.g. introducing parton masses, however, this would only make it unnecessarily complicated~\cite{Ellis:2010rwa,Becher:2016mmh}. Working in $n=4-2\epsilon$ space-time dimensions, and adding all the contributions from the diagrams in Fig.~\ref{fig:Jq}, in $\overline{\rm MS}$ scheme, we have 
\bea
J_q(z, \omega_J) =& g_s^2 \left(\frac{\mu^2e^{\gamma_E}}{4\pi}\right)^{\epsilon} C_F \int \frac{d\ell^+}{2\pi}\frac{1}{\ell^+} \int\frac{d^nq}{(2\pi)^n} \left[4\frac{\ell^+}{q^-} + 2(1-\epsilon) \frac{\ell^+ - q^+}{\omega - q^-}\right]
\nnu
&\times 2\pi\delta(q^+q^- - q_\perp^2) 2\pi\delta\left(\ell^+ - q^+ - \frac{q_\perp^2}{\omega - q^-}\right)
\delta\left(z - \frac{\omega_J}{\omega}\right)
\nnu
&\times
\theta(q^-)\theta(q^+)\theta(\omega-q^-) \theta(\ell^+ - q^+) \Theta_{\rm alg},
\eea
where $\Theta_{\rm alg}$ is the constraint from the jet algorithm and will be discussed separately below for different situations. 

\bef
\includegraphics[width=4.2in]{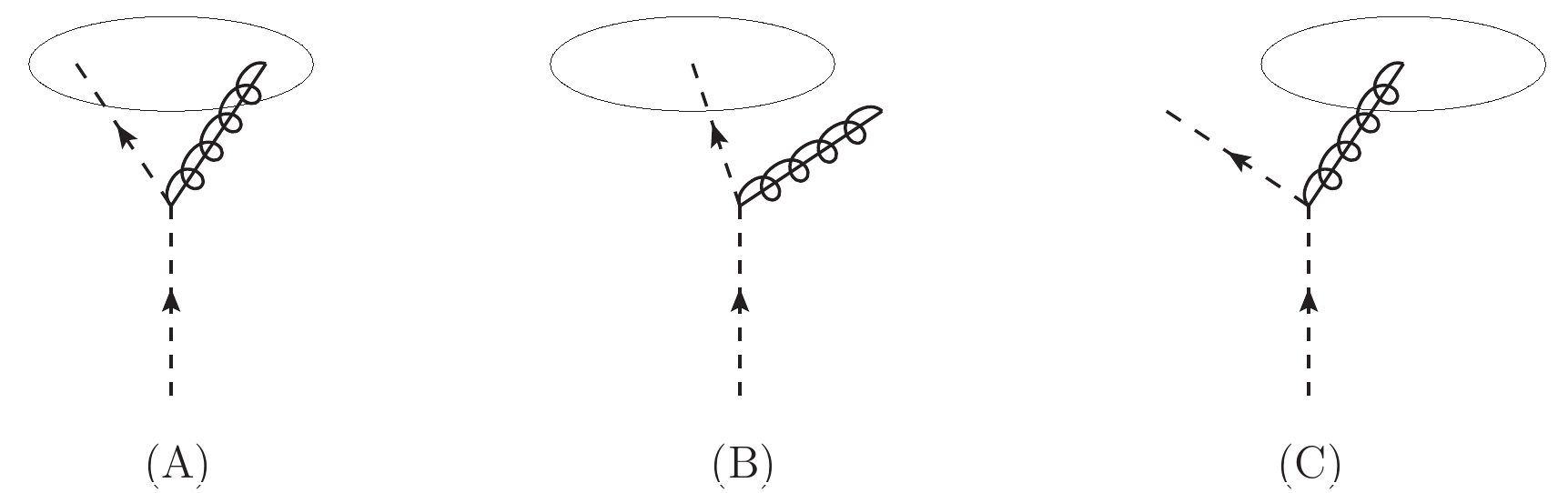}
\caption{Three situations that contribute to the semi-inclusive quark jet function: (A) both quark and gluon are inside the jet, (B) only quark is inside the jet, (C) only gluon is inside the jet.}
\label{fig:configuration}
\eef

There are three situations that we need to consider. We discuss them one by one. 
\begin{enumerate}
\item
Both quark and gluon are inside the jet

The situation is shown in Fig.~\ref{fig:configuration}(A). In this case, the incoming quark energy $\omega$ is the same as the jet energy $\omega_J$, and thus $z=\omega_J/\omega=1$. The constraints for cone and \kt~ algorithms were derived for $e^+e^-$ collisions in \cite{Ellis:2010rwa}. They impose angle restrictions and can be translated to our case as 
\bea
\text{cone:~~} & \Theta_{\rm cone} =  \theta\left(\tan^2\frac{\R}{2} - \frac{q^+}{q^-}\right)  \theta\left(\tan^2\frac{\R}{2} - \frac{\ell^+ - q^+}{\omega - q^-}\right),
\\
\text{\kt:~~} & \Theta_{\text{\kt}} =   \theta\left(\tan^2\frac{\R}{2} - \frac{q^+\omega^2}{q^-\left(\omega - \ell^-\right)^2}\right).
\eea
If we define $x = (\ell - q)^-/\ell^-$, with $q_\perp^2 = q^+q^-$, we can rewrite the above constraints as follows:
\bea
\text{cone:~~} & \Theta_{\rm cone} = \theta\left((1-x)\omega_J \tan\frac{\R}{2} - q_\perp\right) \theta\left(x\omega_J \tan\frac{\R}{2} - q_\perp\right),
\\
\text{\kt:~~} & \Theta_{\text{\kt}} = \theta\left(x(1-x)\omega_J\tan\frac{\R}{2} - q_\perp\right).
\eea

Performing the integration over $\ell^+,~q^+$, we end up with the following expression,
\bea
J_{q\to qg}(z, \omega_J)= \delta(1-z) \frac{\alpha_s}{\pi} \frac{(\mu^2e^{\gamma_E})^{\epsilon}}{\Gamma(1-\epsilon)} \int_0^1 dx \hat P_{qq}(x, \epsilon) \int 
\frac{dq_\perp}{q_\perp^{1+2\epsilon}} \Theta_{\rm alg},
\eea
where the subscript ``$qg$'' represents the situation with both $q$ and $g$ inside the jet, and the function $\hat P_{qq}(x, \epsilon)$ is given by
\bea
\hat P_{qq}(x, \epsilon) = C_F\left[\frac{1+x^2}{1-x} - \epsilon (1-x)\right].
\eea

Implementing the constraints from the jet algorithms, e.g., for \kt~algorithm, we have
\bea
\int  
\frac{dq_\perp}{q_\perp^{1+2\epsilon}} \Theta_{\text{\kt}} = \int_0^{x(1-x)\omega_J\tan\frac{\R}{2}}  \frac{dq_\perp}{q_\perp^{1+2\epsilon}} = - \frac{1}{2\epsilon} \left(\omega_J \tan\frac{\R}{2}\right)^{-2\epsilon} \left(x(1-x)\right)^{-2\epsilon}.
\eea
Further integrating over $x$, and performing the $\epsilon$-expansion, we obtain for the \kt~algorithm,
\bea
J_{q\to qg}(z, \omega_J)  \stackrel{\text{\kt}}{=} \delta(1-z) \frac{\alpha_s}{2\pi} C_F
\left[\frac{1}{\epsilon^2} + \frac{3}{2\epsilon} + \frac{1}{\epsilon} L+ \frac{1}{2}L^2+\frac{3}{2}L + \frac{13}{2} - \frac{3\pi^2}{4}\right],
\eea
where $L$ is defined as
\bea
L = \ln\frac{\mu^2}{\omega_J^2\tan^2\frac{\R}{2}}.
\eea
For the cone-jet algorithm, we can derive the results accordingly. Let us express the results collectively as
\bea
J_{q\to qg}(z, \omega_J) = \delta(1-z) \frac{\alpha_s}{2\pi}
\left[C_F\left(\frac{1}{\epsilon^2} + \frac{3}{2\epsilon} + \frac{1}{\epsilon} L+ \frac{1}{2}L^2+\frac{3}{2}L\right) + d_q^{\rm alg}\right],
\label{eq:qg}
\eea
where the constant term $d_q^{\rm alg}$ depends on the jet algorithm and is given by
\bea
d_q^{\rm cone} &= C_F\left(\frac{7}{2} + 3 \ln2 - \frac{5\pi^2}{12}\right),
\\
d_q^{\rm \text{\kt}} &= C_F\left( \frac{13}{2} - \frac{3\pi^2}{4}\right).
\eea
It is worthwhile to point out that $J_{q\to qg}(z, \omega_J)$ in Eq.~\eqref{eq:qg} is exactly the same as the so-called unmeasured quark jet function in~\cite{Ellis:2010rwa}, multiplied by the factor $\delta(1-z)$. 

\item
Only the quark is inside the jet

The situation is illustrated in Fig.~\ref{fig:configuration}(B). In this case, the final-state quark forms the jet, with a jet energy $\omega_J = (\ell - q)^- = z\, \ell^-$. In other words, only a fraction $z$ of the incoming quark energy $\omega$ is translated into the jet energy. The constraints from the jet algorithms, i.e. the gluon is outside the jet which is composed by the final-state quark only, is the same for cone and \kt~algorithms, and is simply given by
\bea
\Theta_{\rm cone} =  \Theta_{\text{\kt}} =   \theta\left(\frac{q^+\omega^2}{q^-\left(\omega - \ell^-\right)^2} - \tan^2\frac{\R}{2} \right).
\label{eq:qinoriginal}
\eea
Using $\omega_J = (\ell - q)^- = z\, \ell^-$ with $\ell^- = \omega$ and $q_\perp^2 = q^+q^-$, we can rewrite the constraint as
\bea
\Theta_{\rm cone} = \Theta_{\text{\kt}} = \theta\left(q_\perp - (1-z) \omega_J\tan\frac{\R}{2}\right).
\label{eq:quarkin}
\eea

Following the same calculation as above, we obtain
\bea
J_{q\to q(g)}(z, \omega_J) = \frac{\alpha_s}{\pi}  \frac{(\mu^2e^{\gamma_E})^{\epsilon}}{\Gamma(1-\epsilon)} \hat P_{qq}(z, \epsilon) 
\int \frac{dq_\perp}{q_\perp^{1+2\epsilon}} \Theta_{\rm alg},
\eea
where the subscript ``$q(g)$'' represents the situation with only $q$ inside and $g$ outside the jet. The jet algorithm then leads to the following constraint
\bea
\int \frac{dq_\perp}{q_\perp^{1+2\epsilon}} \Theta_{\rm alg} = \int_{(1-z)\omega_J\tan\frac{\R}{2}}^{\infty} \frac{dq_\perp}{q_\perp^{1+2\epsilon}} = \frac{1}{2\epsilon} \left(\omega_J \tan\frac{\R}{2}\right)^{-2\epsilon} \left(1-z\right)^{-2\epsilon}.
\eea
Performing the $\epsilon$-expansion, we thus have
\bea
\label{eq:q(g)}
J_{q\to q(g)}(z, \omega_J) =& \frac{\alpha_s}{2\pi} C_F \delta(1-z)\left[-\frac{1}{\epsilon^2} -\frac{1}{\epsilon} L - \frac{1}{2}L^2 + \frac{\pi^2}{12}\right]
\\
&+\frac{\alpha_s}{2\pi} C_F \left[\left(\frac{1}{\epsilon}+L\right) \frac{1+z^2}{(1-z)_+} -2(1+z^2)\left(\frac{\ln(1-z)}{1-z}\right)_+ - (1-z)\right].
\nonumber
\eea

\item
Only the gluon is inside the jet

The situation is illustrated in Fig.~\ref{fig:configuration}(C). In this case, the final-state gluon forms the jet, with a jet energy $\omega_J = q^- = z\, \ell^-$. It is easy to be convinced that the constraint from the jet algorithms are again given by Eq.~\eqref{eq:qinoriginal} or Eq.~\eqref{eq:quarkin}. 

The calculation is very similar to the case where the quark is inside the jet, and we obtain
\bea
J_{q\to (q)g}(z, \omega_J) = \frac{\alpha_s}{2\pi}\left(\frac{1}{\epsilon} + L\right) P_{gq}(z) - \frac{\alpha_s}{2\pi} \Big[P_{gq}(z) 2 \ln(1-z) + C_F z \Big],
\label{eq:g(q)}
\eea
where the subscript ``$(q)g$'' represents the situation with $g$ inside and $q$ outside the jet. It is worthwhile to emphasize that the situations 2 and 3 do not have a jet algorithm-dependence, simply because only one particle forms the jet. 

\end{enumerate}

Ellis et al. also considered the above three situations in their seminal work~\cite{Ellis:2010rwa}, hence it is instructive to compare to their results. In~\cite{Ellis:2010rwa}, the authors place an energy cut $\Lambda$ on the total energy outside of the observed jets to ensure that the jet algorithm does not find more than $N$ jets. In such a case of {\it exclusive} jet production, the parton outside the jet should have energy less than $\Lambda$. It was shown carefully in~\cite{Ellis:2010rwa} that the contributions from the above situations 2 and 3 (i.e. only one parton is inside the jet) are power suppressed by ${\mathcal O}(\Lambda/\omega)$. However, this is not the situation we consider in our current paper. Here, we have in mind the {\it inclusive} jet production, and we do not place any constraint on the energy of the parton outside the jet. As long as the jet energy $\omega_J$ is large enough to be observed as a jet following the experimental kinematic cuts, it will be identified as a jet. In this case, the contributions from 2 and 3 are not power suppressed, as can be clearly seen from the expressions above. 

Summing the above three contributions, we obtain the full expression for the semi-inclusive quark jet function
\bea
J_q^{(1)}(z, \omega_J) = & J_{q\to qg}(z, \omega_J) + J_{q\to q(g)}(z, \omega_J) + J_{q\to (q)g}(z, \omega_J)
\label{eq:J1qterms}
\\
=& \frac{\alpha_s}{2\pi} \left(\frac{1}{\epsilon} + L\right) 
\Big[P_{qq}(z) + P_{gq}(z)\Big] 
\nnu
&-\frac{\alpha_s}{2\pi} \Bigg\{ C_F\left[2\left(1+z^2\right)\left(\frac{\ln(1-z)}{1-z}\right)_{+} + (1-z) \right] - \delta(1-z) d_J^{q,{\rm alg}}
\nnu
&+P_{gq}(z) 2\ln\left(1-z\right) + C_F z \Bigg\},
\label{eq:J1q}
\eea
where the superscript ``(1)'' represents the NLO ${\mathcal O}(\alpha_s)$ result, $P_{qq}(z)$ and $P_{gq}(z)$ are the standard Altarelli-Parisi splitting functions,
\bea
P_{qq}(z) &= C_F \left[\frac{1+z^2}{(1-z)_+} + \frac{3}{2}\delta(1-z) \right],
\label{eq:Pqq}
\\
P_{gq}(z) &= C_F \frac{1+(1-z)^2}{z}.
\label{eq:Pgq}
\eea
On the other hand, the constant term $d_J^{q,{\rm alg}}$ depends on the jet algorithm, and they are related to $d_q^{\rm alg}$ in Eq.~\eqref{eq:qg} by
\bea
d_J^{q,\rm alg} = d_q^{\rm alg} + C_F \frac{\pi^2}{12}, 
\eea
where the second term comes from the constant $\delta(1-z)$-piece in Eq.~\eqref{eq:q(g)}. It might be instructive to point out that this second term actually corresponds to the same $\pi^2$-constant term of the single hemisphere soft function~\cite{Kolodrubetz:2016dzb}, and such a fact thus demonstrates the consistency with the exclusive limit at $z\to 1$~\footnote{We thank P.~Pietrulewicz, I.~Stewart, F.~ Tackmann, and W. Waalewijn for pointing this out.}. The constant terms $d_J^{q,\rm alg}$ have the following explicit expressions,
\bea
d_J^{q,\rm cone} &= C_F\left(\frac{7}{2} + 3 \ln2 - \frac{\pi^2}{3}\right),
\\
d_J^{q,\rm \text{\kt}} &= C_F\left( \frac{13}{2} - \frac{2\pi^2}{3}\right).
\eea
Adding LO to NLO results, we obtain the full result for the semi-inclusive quark jet function,
\bea
J_q(z, \omega_J) = J_q^{(0)}(z, \omega_J)+J_q^{(1)}(z, \omega_J).
\eea
It is very interesting to point out that although the contribution with both $q$ and $g$ inside the jet, 
$J_q(z, \omega_J)|_{qg}$, contains a double pole $1/\epsilon^2$ (correspondingly the double logarithm $L^2$), such double poles and $L^2$ cancel out between $J_{q\to qg}(z, \omega_J)$ and $J_{q\to q(g)}(z, \omega_J)$. We are thus left with only a single pole $1/\epsilon$ and the single logarithm $L$ for $J_q(z, \omega_J)$. Such a difference is the main reason why the unmeasured jet function $J_q(\omega_J)$ widely studied in SCET (see, e.g.,~\cite{Ellis:2010rwa}) will follow RG evolution equations different from our semi-inclusive jet functions $J_q(z, \omega_J)$, as we will demonstrate below. 

\subsection{The semi-inclusive gluon jet function}
\bef
\includegraphics[width=\textwidth]{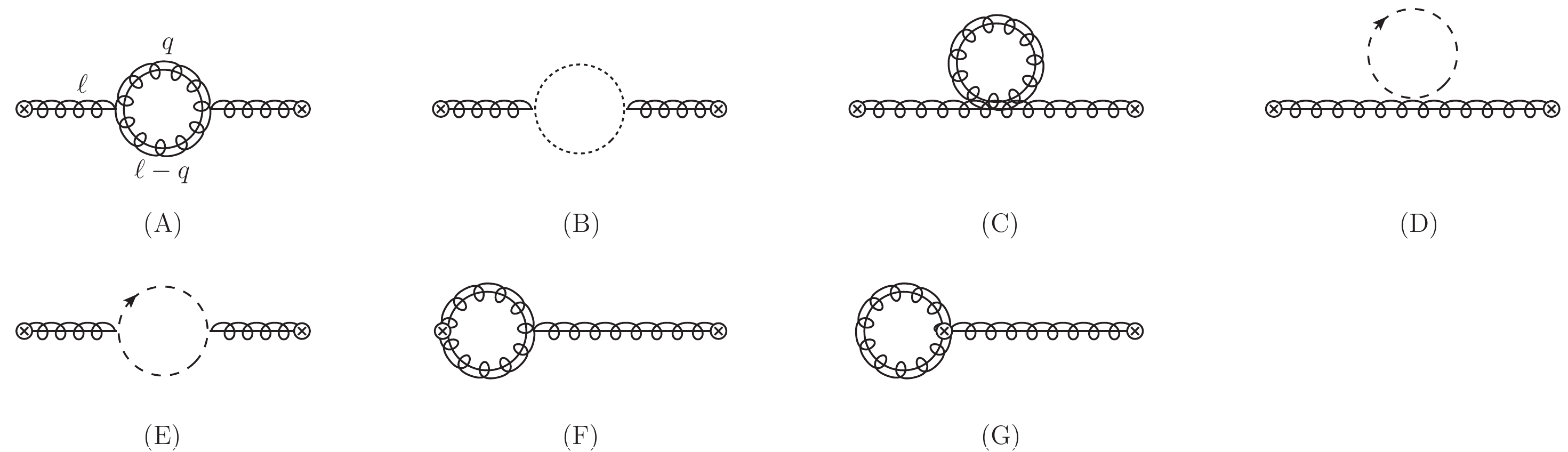}
\caption{Feynman diagrams that contribute to the semi-inclusive gluon jet function $J_g(z, \omega_J)$. The gluon that initiates the jet has momentum $\ell = (\ell^- = \omega, \ell^+, 0_\perp)$, with $\omega = \omega_J/z$ and $\omega_J$ the jet energy. The dotted loop in (B) is the ghost loop, while the dashed loop in (D) and (E) are collinear quark loops, the mirror diagrams of (F) and (G) are not shown here but are included in the calculations.}
\label{fig:Jg}
\eef

Likewise, we can compute the semi-inclusive gluon jet function $J_g(z, \omega_J)$. The relevant Feynman diagrams are give in Fig.~\ref{fig:Jg}. It also receives three contributions just like $J_q(z, \omega_J)$. When both final-state partons are inside the jet, we have
\bea
J_{g\to gg+q\bar q}(z, \omega_J) \equiv & J_{g\to gg}(z, \omega_J) + J_{g\to q\bar q}(z, \omega_J)
\\
=&\delta(1-z) \frac{\alpha_s}{\pi} \frac{(\mu^2e^{\gamma_E})^{\epsilon}}{\Gamma(1-\epsilon)} \int_0^1 dx 
\left[\hat P_{gg}(x, \epsilon) + 2 n_f \hat P_{qg}(x, \epsilon) \right]
\int \frac{dq_\perp}{q_\perp^{1+2\epsilon}} \Theta_{\rm alg}.
\nonumber
\eea
Here, $J_{g\to gg}$ represents the contribution from $g\to gg$ with both gluons inside the jet, and it is given by the term $\propto \hat P_{gg}(x, \epsilon)$. On the other hand, $J_{g\to q\bar q}$ stands for the contribution from $g\to q\bar q$ with both quark and anti-quark inside the jet, and it is given by the term $\propto \hat P_{qg}(x, \epsilon)$. Here $\hat P_{qg}(x, \epsilon)$ and $\hat P_{gg}(x, \epsilon)$ are given by
\bea
\hat P_{qg}(x, \epsilon) =& T_F\left[1-\frac{2x(1-x)}{1-\epsilon}\right],
\\
\hat P_{gg}(x, \epsilon) =& 2C_A\left[\frac{x}{1-x} + \frac{1-x}{x} + x(1-x)\right].
\eea
After taking into account the constraint from the jet algorithm, and completing the integration and $\epsilon$-expansion, we obtain
\bea
J_{g\to gg+q\bar q}(z, \omega_J) =& \delta(1-z) \frac{\alpha_s}{2\pi} 
\left(\frac{C_A}{\epsilon^2} + \frac{\beta_0}{2\epsilon} + \frac{C_A}{\epsilon} L\right.
+\left.\frac{C_A}{2}L^2 + \frac{\beta_0}{2} L + d_g^{\rm alg}\right),
\label{eq:Jgg}
\eea
where $\beta_0$ is the lowest order coefficient of the QCD $\beta$ function,
\bea
\beta_0 = \frac{11}{3}C_A - \frac{4}{3} T_F n_f,
\eea
and the constant terms $d_g^{\rm alg}$ have the following expressions
\bea
d_g^{\rm cone} &=C_A \left(\frac{137}{36} + \frac{11}{3}\ln2 - \frac{5\pi^2}{12} \right) - T_F n_f \left(\frac{23}{18}+\frac{4}{3}\ln2\right),
\label{eq:dgcone}
\\
d_g^{\rm \text{\kt}} & = C_A \left(\frac{67}{9} - \frac{3\pi^2}{4}\right) - T_Fn_f \left(\frac{23}{9}\right).
\label{eq:dgkt}
\eea
On the other hand, when one of the partons is outside the jet, we have
\bea
J_{g\to g(g)}(z, \omega_J) &= J_{g\to (g)g}(z, \omega_J),
\label{eq:Jg(g)s}
\\
J_{g\to q(\bar q)}(z, \omega_J) &= J_{g\to (q)\bar q}(z, \omega_J),
\label{eq:Jq(qb)s}
\eea
where the subscript ``$g(g)$'' on the left-hand side means that only the gluon $g$ with momentum $\ell-q$ is inside the jet, while ``$(g)g$'' on the right-hand side represents that only the gluon $g$ with momentum $q$ is inside the jet. They are symmetric, and thus give the same results. Similar is the case of $g\to q(\bar q)$ and $g\to (q)\bar q$. To simplify the notation, in the rest of the paper, we use $J_{g\to g(g) + q(\bar q)}(z, \omega_J)$ to represent the sum of both cases. The result is given by
\bea
J_{g\to g(g) + q(\bar q)}(z, \omega_J) = 2\,\frac{\alpha_s}{\pi} \frac{(\mu^2e^{\gamma_E})^{\epsilon}}{\Gamma(1-\epsilon)}  
\left[\hat P_{gg}(z, \epsilon) + 2 n_f \hat P_{qg}(z, \epsilon) \right]
\int \frac{dq_\perp}{q_\perp^{1+2\epsilon}} \Theta_{\rm alg},
\eea
where the factor of ``2'' on the right hand side is reflecting the identities in Eqs.~\eqref{eq:Jg(g)s} and \eqref{eq:Jq(qb)s}. With the constraint from the jet algorithm in Eq.~\eqref{eq:quarkin}, we can integrate over $q_\perp$ and perform the $\epsilon$-expansion. The final result is given by
\bea
J_{g\to g(g) + q(\bar q)}(z, \omega_J) =& \frac{\alpha_s}{2\pi} \delta(1-z) \left(
-\frac{C_A}{\epsilon^2} - \frac{\beta_0}{2\epsilon} - \frac{C_A}{\epsilon} L - \frac{C_A}{2}L^2 -\frac{\beta_0}{2}L+ \frac{\pi^2}{12}\right)
\nnu
&+\frac{\alpha_s}{2\pi}\left(\frac{1}{\epsilon}+L\right)\Big[P_{gg}(z) + 2n_f P_{qg}(z) \Big]
\nnu
&- \frac{\alpha_s}{2\pi} \Bigg[ \frac{4C_A (1-z+z^2)^2}{z} \left(\frac{\ln(1-z)}{1-z}\right)_{+} 
\nnu
& + 4n_f\Big(P_{qg}(z)\ln(1-z) + T_F z(1-z)\Big) \Bigg],
\label{eq:Jg(g)}
\eea
where $P_{gg}(z)$ and $P_{qg}(z)$ are the standard splitting functions with the expressions,
\bea
P_{gg}(z) &= 2C_A \left[\frac{z}{(1-z)_+} + \frac{1-z}{z} +z(1-z) \right] + \frac{\beta_0}{2} \delta(1-z),
\label{eq:Pgg}
\\
P_{qg}(z) &= T_F\left[z^2+(1-z)^2\right].
\label{eq:Pqg}
\eea
Adding the contributions from Eqs.~\eqref{eq:Jgg} and \eqref{eq:Jg(g)} together, we obtain the following expression for the semi-inclusive gluon jet function $J_g(z, \omega_J)$ at NLO,
\bea
J_g^{(1)}(z, \omega_J) =&  J_{g\to gg + q\bar q}(z, \omega_J)  + J_{g\to g(g) + q(\bar q)}(z, \omega_J) 
\nnu
=&   \frac{\alpha_s}{2\pi} \left(\frac{1}{\epsilon} + L\right) 
\Big[P_{gg}(z) + 2n_f P_{qg}(z)\Big]
\nnu
&- \frac{\alpha_s}{2\pi} \Bigg[ \frac{4C_A (1-z+z^2)^2}{z} \left(\frac{\ln(1-z)}{1-z}\right)_{+}  - \delta(1-z) d_J^{g,{\rm alg}}
\nnu
& + 4n_f\Big(P_{qg}(z)\ln(1-z) + T_F z(1-z)\Big) \Bigg],
\label{eq:J1g}
\eea
where again $d_J^{g,{\rm alg}}$ is related to $d_g^{\rm alg}$ as follows
\bea
d_J^{g,{\rm alg}} = d_g^{\rm alg} + C_A\frac{\pi^2}{12},
\eea
with $d_g^{\rm alg}$ given in Eqs.~\eqref{eq:dgcone} and \eqref{eq:dgkt}. For later convenience and completeness, we give them here:
\bea
d_J^{g,\rm cone} &=C_A \left(\frac{137}{36} + \frac{11}{3}\ln2 - \frac{\pi^2}{3} \right) - T_F n_f \left(\frac{23}{18}+\frac{4}{3}\ln2\right),
\\
d_J^{g,\rm \text{\kt}} & = C_A \left(\frac{67}{9} - \frac{2\pi^2}{3}\right) - T_Fn_f \left(\frac{23}{9}\right).
\eea
Again, we find that all double pole $1/\epsilon^2$ and the double logarithms $L^2$ cancel between the above contributions, and we are left with only a single pole $1/\epsilon$ and a single logarithm~$L$. 

\subsection{RG evolution}
We will now discuss the renormalization of the above semi-inclusive jet functions. The renormalized semi-inclusive jet functions are defined through
\bea
J_{i, \rm bare}(z, \omega_J) = \sum_j \int_z^1 \frac{dz'}{z'} Z_{ij}\left(\frac{z}{z'}, \mu \right) J_j(z', \omega_J, \mu),
\label{eq:bare}
\eea
with $Z_{ij}$ the renormalization matrix. The renormalization-group equation for the renormalized semi-inclusive jet functions $J_i(z, \omega_J, \mu)$ will thus follow from Eq.~\eqref{eq:bare},
\bea
\mu\frac{d}{d\mu} J_i(z, \omega_J, \mu) = \sum_j \int_z^1 \frac{dz'}{z'} \gamma_{ij}^J \left(\frac{z}{z'}, \mu \right) J_j(z', \omega_J, \mu),
\eea
with anomalous dimension $\gamma_{ij}^J$ given by 
\bea
\gamma_{ij}^J (z, \mu) = - \sum_k \int_z^1 
\frac{dz'}{z'} \left(Z\right)^{-1}_{ik} \left(\frac{z}{z'}, \mu\right) 
\mu \frac{d}{d\mu} Z_{kj}(z', \mu).
\label{eq:gamma_ij}
\eea
Here, the inverse of the renormalization factor $\left(Z\right)^{-1}_{ik}$ is defined through
\bea
\sum_k \int_z^1 \frac{dz'}{z'} \left(Z\right)^{-1}_{ik}\left(\frac{z}{z'}, \mu\right)  Z_{kj}(z', \mu)
=\delta_{ij} \delta(1-z).
\eea

The lowest order renormalization factors $Z_{ij}^{(0)}$ can be trivially determined,
\bea
Z_{ij}^{(0)}(z, \mu) = \delta_{ij} \delta(1-z).
\eea
On the other hand, the one-loop renormalization factors $Z_{ij}^{(1)}$ can be extracted from our one-loop results presented in last section, Eqs.~\eqref{eq:J1q} and \eqref{eq:J1g}. We obtain to NLO,
\bea
Z_{ij}(z, \mu) = \delta_{ij} \delta(1-z) + \frac{\alpha_s(\mu)}{2\pi} \left(\frac{1}{\epsilon}\right) P_{ji}(z),
\eea
where $P_{ji}(z)$ are the standard splitting functions as given in Eqs.~\eqref{eq:Pqq}, \eqref{eq:Pgq}, \eqref{eq:Pgg}, \eqref{eq:Pqg}. Thus, the renormalized semi-inclusive jet functions at NLO have the following expressions
\bea
J_i(z, \omega_J, \mu) = J^{(0)}_i(z, \omega_J, \mu) + J^{(1)}_i(z, \omega_J, \mu),
\eea
where $J^{(0)}_i(z, \omega_J, \mu) = \delta(1-z)$, and 
\bea
J_q^{(1)}(z, \omega_J, \mu) =& \frac{\alpha_s}{2\pi} L \Big[P_{qq}(z) + P_{gq}(z)\Big] 
-\frac{\alpha_s}{2\pi} \Bigg\{ C_F\left[2\left(1+z^2\right)\left(\frac{\ln(1-z)}{1-z}\right)_{+} + (1-z) \right] 
\nnu
& - \delta(1-z) d_J^{q,{\rm alg}} +P_{gq}(z) 2\ln\left(1-z\right) + C_F z \Bigg\},
\label{eq:ren-Jq}
\\
J_g^{(1)}(z, \omega_J, \mu)=&   \frac{\alpha_s}{2\pi}  L \Big[P_{gg}(z) + 2n_f P_{qg}(z)\Big]
- \frac{\alpha_s}{2\pi} \Bigg[ \frac{4C_A (1-z+z^2)^2}{z} \left(\frac{\ln(1-z)}{1-z}\right)_{+}  
\nnu
& - \delta(1-z) d_J^{g,{\rm alg}} + 4n_f\Big(P_{qg}(z)\ln(1-z) + T_F z(1-z)\Big) \Bigg].
\label{eq:ren-Jg}
\eea
It is interesting to point out that the above renormalized semi-inclusive jet functions are exactly the same as those found through conventional NLO calculations for single inclusive jet cross section, see, \cite{Jager:2004jh,Mukherjee:2012uz,Kaufmann:2015hma}. 

On the other hand, from Eq.~\eqref{eq:gamma_ij} we obtain the anomalous dimensions of the semi-inclusive jet functions
\bea
\gamma_{ij}^J(z, \mu) =  \frac{\alpha_s(\mu)}{\pi}  P_{ji}(z). 
\eea
We thus have the following RG evolution for $J_{q/g}(z, \omega_J, \mu)$
\bea
\mu \frac{d}{d\mu} J_i(z, \omega_J, \mu) = \frac{\alpha_s(\mu)}{\pi} \sum_j \int_z^1  \frac{dz'}{z'} P_{ji}\left(\frac{z}{z'}, \mu \right) J_j(z', \omega_J, \mu).
\label{eq:evo}
\eea
In other words, they are exactly the same as the usual timelike DGLAP evolution equations for standard fragmentation functions $D_i^h(z, \mu)$. 

It is instructive to point out that from the NLO expressions in Eqs.~\eqref{eq:ren-Jq} and \eqref{eq:ren-Jg}, the natural scale for $J_i(z, \omega_J, \mu)$ is given by
\bea
\mu \sim \omega_J \tan\frac{\R}{2} \equiv \mu_J,
\eea
at which the large logarithmic terms $\sim L$ are eliminated. Realizing that $\omega_J = 2 p_T \cosh \eta$, we have
\bea
\mu_J = \omega_J \tan\frac{\R}{2} = \left(2 p_T \cosh \eta \right)\tan\left(\frac{R}{2\cosh \eta}\right) \approx p_T R,
\eea
where we have used Eq.~\eqref{eq:algorithm} for the expression of $\R$, and $\tan (x) \approx x$ for small $x$. Thus, solving the above evolution equations from the scale $\mu_J \sim p_T R$ to a higher scale $\mu\sim p_T$, we  naturally resum the logarithms of the form $(\as \ln R )^n$, which can be large for small $R$. For later convenience, let us denote the natural scale of the semi-inclusive jet functions as 
\bea
p_{TR} \equiv p_T \, R.
\label{eq:pTR}
\eea
We will demonstrate such a small jet radius resummation for single inclusive jet production below. 

\subsection{Small jet radius resummation}
\label{sec:solution}
Following Eq.~(\ref{eq:evo}), the timelike DGLAP evolution equations for the semi-inclusive jet function can be cast into the following form
\be\label{eq:DGLAP2}
\frac{d}{d \log \mu^2}
\begin{pmatrix}
J_S (z,\omega_J,\mu) \\ J_g(z,\omega_J,\mu)
\end{pmatrix}
=
\frac{\alpha_s(\mu)}{2 \pi}
\begin{pmatrix}
P_{qq}(z) & ~2 N_f P_{gq}(z) \\
P_{qg}(z) & ~P_{gg}(z)
\end{pmatrix}
\otimes
\begin{pmatrix}
J_S (z,\omega_J,\mu)\\ J_g(z,\omega_J,\mu)
\end{pmatrix},
\ee
where $\otimes$ denotes the usual convolution integral defined as
\be
(f\otimes g)(z)=\int_z^1\f{dz'}{z'} f(z')g(z/z') \, .
\ee
The function $J_S(z,\omega_J,\mu)$ in~(\ref{eq:DGLAP2}) is the singlet semi-inclusive jet function given by the sum over all quark and anti-quark flavors
\be
J_S(z,\omega_J,\mu)=\sum_{q,\bar q}J_q(z,\omega_J,\mu)=2 N_f J_q(z,\omega_J,\mu) \, .
\ee
Since the semi-inclusive jet function is the same for all quarks and anti-quarks, we do not need to consider separate non-singlet evolutions. 

The initial conditions for the evolution equations at the scale $\mu_J$ involve delta functions and distributions. We deal with this problem by solving the evolution equations in Mellin moment space following the method outlined in~\cite{Vogt:2004ns}. The Mellin moments of any $z$-dependent function are defined as
\be
f(N)=\int_0^1 dz\, z^{N-1} f(z) \, .
\ee
Note that the delta functions and ``plus'' distributions turn into simple functions in Mellin moment space. After performing the evolution in Mellin space from scale $\mu_J$ to any scale $\mu$, we take the Mellin inverse transformation in order to obtain the corresponding 
semi-inclusive jet functions in $z$ space, $J_{S,\,g}(z,\omega_J, \mu)$. An important advantage when formulating the solution of the DGLAP evolution equations in Mellin space is that the convolution structure in~(\ref{eq:DGLAP2}) turns into simple products. Schematically, one has
\be
(f\otimes g)(N)=f(N)\, g(N) \, .
\ee
We can write down the solution of the DGLAP equations in Mellin space for an evolution from scale $\mu_J$ to $\mu$ as~\cite{Vogt:2004ns}
\be\label{eq:DGLAP3}
\begin{pmatrix}
J_S (N,\omega_J,\mu) 
\\ J_g (N,\omega_J,\mu)
\end{pmatrix}
=
\left[
e_+(N)
\left(\frac{\alpha_s(\mu)}{\alpha_s(\mu_J)} \right)^{-r_-(N)}
+ e_-(N)
\left(\frac{\alpha_s(\mu)}{\alpha_s(\mu_J)} \right)^{-r_+(N)}
\right]
\begin{pmatrix}
 J_S (N,\omega_J,\mu_J) \\ J_g (N,\omega_J,\mu_J)
\end{pmatrix} ,
\ee
where $r_+(N)$ and $r_-(N)$ denote the larger and smaller eigenvalue of the leading-order singlet evolution matrix, see~(\ref{eq:DGLAP2}),
\be
r_{\pm}(N)=\f{1}{2\beta_0}\left[P_{qq}(N)+P_{gg}(N)\pm\sqrt{\left(P_{qq}(N)-P_{gg}(N)\right)^2 +4 P_{qg}(N) P_{gq}(N)} \right] \, .
\ee
The projector matrices $e_{\pm}(N)$ in~(\ref{eq:DGLAP3}) are defined as
\be
e_{\pm}(N)=\f{1}{r_\pm(N) - r_\mp(N)}
\begin{pmatrix}
P_{qq}(N)-r_\mp(N) & ~2 N_f P_{gq}(N) \\
P_{qg}(N) & ~P_{gg}(N)-r_\mp(N)
\end{pmatrix}
\, .
\ee
The evolved semi-inclusive jet functions in $z$-space are eventually obtained by performing a Mellin inverse transformation
\be\label{eq:MellinInverse}
J_{S,g}(z,\omega_J,\mu) = \f{1}{2\pi i} \int_{{\cal C}_N} dN\, z^{-N} J_{S,g}(N,\omega_J,\mu)\, ,
\ee
where the contour in the complex $N$ plane is chosen to the right of all the poles in $J_{S,g}(N,\omega_J,\mu)$.

Our evolution code is a modified version of the evolution code for fragmentation functions presented in~\cite{Anderle:2015lqa}, which in turn is based on the {\sc Pegasus} evolution package for PDFs~\cite{Vogt:2004ns}. The evolution codes of~\cite{Anderle:2015lqa,Vogt:2004ns} can be used to perform an evolution at NNLO. Here we only need a LO evolution instead. However, for the purpose of this work, we had to increase the numerical precision in the region of $z\to 1$. PDFs and FFs fall off as $\sim(1-z)^\alpha$ for $z\to 1$, where $\alpha$ is typically in the range of $\alpha=3-8$. Instead, here we have to handle distributions at the initial scale $\mu_J$ which are divergent for $z\to 1$. We deal with this divergence by adopting a prescription developed in~\cite{Bodwin:2015iua}, as discussed below.

\bef
\includegraphics[width=4.7in]{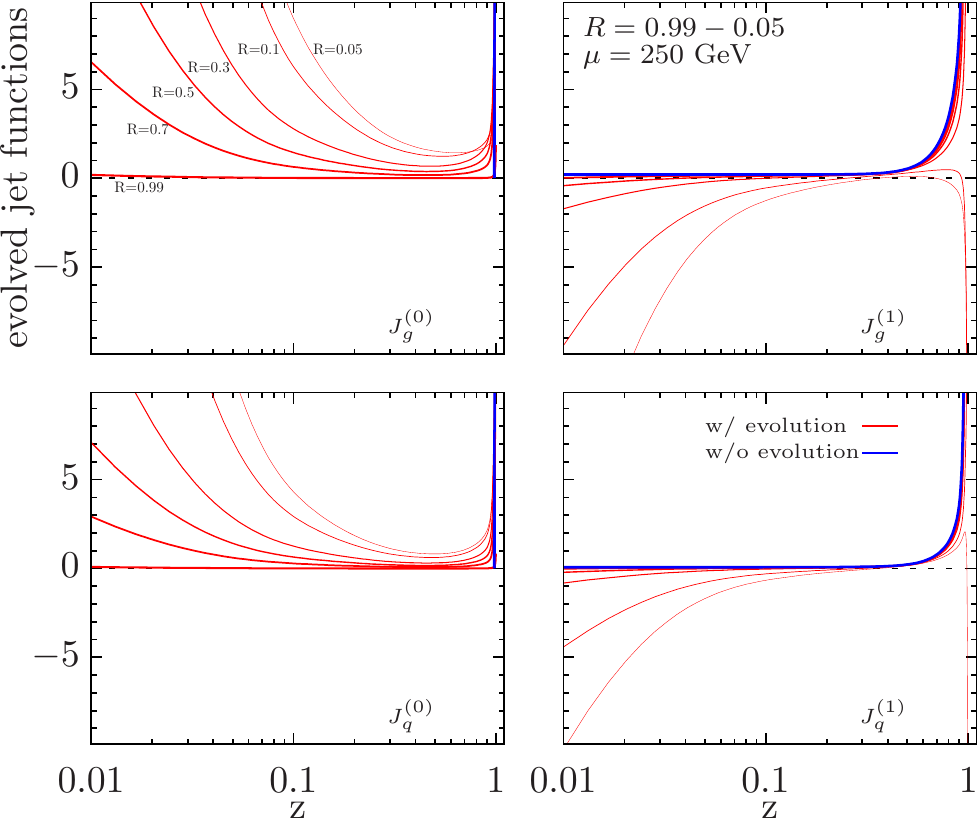}
\caption{The semi-inclusive jet function with evolution (red) and without evolution (blue) for several values of the jet radius parameter $R=0.99,\, 0.7,\, 0.5,\, 0.3,\, 0.1,\, 0.05$. Using the DGLAP evolution equations, the semi-inclusive jet function is evolved to a final scale of $\mu=250$~GeV. In order to perform the correct matching to NLO, we need to perform the evolution of the LO and NLO jet functions separately for both quarks $J_q^{(0),(1)}$ and for gluons $J_g^{(0),(1)}$ as shown in the four panels. Note that the initial condition for the evolution of the LO jet function is given by a delta function which is illustrated in the left two panels by a blue straight line.}
\label{fig:Jqevo}
\eef

Fig.~\ref{fig:Jqevo} shows the evolved (red) and unevolved (blue) jet functions $J_{q,g}(z,\omega_J, \mu)$. As an example, we choose several different values of the jet parameter in the range of $R=0.05-0.99$ and a final scale for the evolution of $\mu=250$~GeV, while we set the initial evolution scale $\mu_J = \mu\, R$ to eliminate the logarithm $L$ in the fixed-order expressions for $J_{q,g}(z,\omega_J, \mu)$. Since the DGLAP equations are {\it linear} evolution equations, the evolution of the sum $\left(J_i^{(0)}+J_i^{(1)}\right)$ will be equal to the sum of the individually evolved $J_i^{(0)}$ and $J_i^{(1)}$. Here we present the evolved $J_i^{(0)}$ and $J_i^{(1)}$ separately for later convenience. In the left two panels, the leading-order jet functions $J_{q,g}^{(0)}(z,\omega_J,\mu)$ are shown. In this case, the initial condition for the evolution is simply given by a delta function $\delta(1-z)$, as illustrated in blue at $z=1$. We note that a longer evolution, i.e. a lower starting scale due to a smaller value of $R$, leads to an increase at small-$z$ as it is expected for an evolution to larger scales. One also notices that the evolution for the gluon is stronger than for the quark semi-inclusive jet function. In the two panels on the right side of Fig.~\ref{fig:Jqevo}, we show the evolution of the ${\cal O}(\as)$ correction for the semi-inclusive jet function at NLO, $J_{q,g}^{(1)}(z,\omega_J, \mu)$. Both initial conditions $J_{q,g}^{(1)}(z,\omega_J, \mu)$ are also divergent at $z=1$ since they contain distributions. Note that in this case, the evolution leads to a decrease both at small- and large-$z$. A sufficiently long evolution can turn the evolved functions negative for both small- and large-$z$.

\section{Application: $e^+e^-\to {\rm jet} X$}
\label{sec:ee}
In this section we consider single inclusive jet production in $e^+ e^-$ collisions, $e^+  e^-\to {\rm jet}X$. We demonstrate to the next-to-leading order that the short distance hard functions for single jet production are the same as those for single hadron production, $e^+  e^-\to hX$, with only the standard fragmentation functions $D_{i}^h(z, \mu)$ replaced by the semi-inclusive jet functions $J_i(z, \omega_J, \mu)$.

\subsection{Factorized form}
To be specific, we study single inclusive jet production, as well as single inclusive hadron production for comparison,
\bea
e^+(k_1) + e^-(k_2) &\to {\rm jet}(p) + X(p_X),
\\
e^+(k_1) + e^-(k_2) &\to h(p) + X(p_X),
\eea
where $X$ denotes all other final-state particles besides the measured jet or hadron, with momentum $p_X$. For simplicity, we assume $e^+e^-$ annihilates into a virtual photon to demonstrate our derivation. The virtual photon has four-momentum $q = k_1+k_2$ with the center-of-mass energy $\sqrt{s} \equiv Q = \sqrt{q^2}$. We are interested in the region where $p_X^2\sim Q^2$, for which a standard collinear factorization theorem has been proven for single inclusive hadron production in the traditional QCD methods, see, e.g., Refs.~\cite{Ellis:1978sf,Ellis:1978ty,Collins:1981ta,Collins:1989gx}. Here, we will first review the same factorization formalism within SCET for single hadron production~\cite{Fickinger:2016rfd} and  then generalize the factorization formalism to single jet production. We find that the factorized forms are given by
\bea
\frac{d\sigma^h}{dp_T d\eta} &= \sum_{c=q,g} \int \frac{dz_c}{z_c} 
H_{e^+e^-\to c}\left(\hat p, \mu \right)
D_{c}^{h}(z_c, \mu),
\label{eq:hadron}
\\
\frac{d\sigma^{\rm jet}}{dp_T d\eta} &= \sum_{c=q,g} \int \frac{dz_c}{z_c} 
H_{e^+e^-\to c}\left(\hat p, \mu \right) J_{c}(z_c, \omega_J, \mu),
\label{eq:jet}
\eea
where $\hat p = p/z_c$ is the four-momentum for the parton that fragments into the final-state hadron $h$ (or that initiates the jet), $\eta$ and $p_T$ are the rapidity and transverse momentum of the hadron (or jet) in the center-of-mass frame of the incoming leptons, and the jet energy $\omega_J = 2 p_T \cosh \eta$. Here in Eqs.~\eqref{eq:hadron} and \eqref{eq:jet}, we use exactly the same short-distance hard functions $H_{e^+e^-\to c}\left(\hat p, \mu \right)$, since we will demonstrate that they are the same below. We choose the cross sections under investigation in Eqs.~\eqref{eq:hadron} and \eqref{eq:jet} to be differential in $p_T$ and $\eta$~\footnote{This is different from the conventional set-up where one usually computes the cross sections as a function of the hadron/jet energy. Nevertheless, there are experimental jet measurements based on our set-up, see, e.g. \cite{Akers:1994wj}.}, because we want to easily generalize the formalism from $e^+e^-$ to $pp$ collisions in the next section. 

We start with the invariant amplitude $M$ for the process to produce a hadron/jet. The invariant amplitude $M$ can be written as
\bea
M_h &= \bar{v}(k_1, \lambda_1) \gamma_{\mu} u(k_2, \lambda_2) \frac{e^2}{Q^2} \langle hX| J^\mu(0) |0\rangle,
\\
M_{\rm jet} &= \bar{v}(k_1, \lambda_1) \gamma_{\mu} u(k_2, \lambda_2) \frac{e^2}{Q^2} \langle J X| 
J^\mu(0) |0\rangle,
\eea
where the subscript $h$ (jet) represents the hadron (jet) production, and the current $J_\mu(0)$ on the hadronic side is 
\bea
J^{\mu}(0) = \bar \psi_q(0) \gamma^\mu \psi_q(0).
\eea
After taking into account the averaging over the incoming polarizations, and at the same time including the final-state phase space, the cross section can be eventually written as
\bea
\frac{d\sigma^h}{dy dp_T} &= \frac{\alpha_{\rm em}^2 p_T}{2Q^6} L_{\mu\nu} W^{\mu\nu}_h,
\label{eq:h1}
\\
\frac{d\sigma^{\rm jet}}{dy dp_T} &= \frac{\alpha_{\rm em}^2 p_T}{2Q^6} L_{\mu\nu} W^{\mu\nu}_{\rm jet},
\label{eq:j1}
\eea
where the leptonic tensor $L_{\mu\nu}$ has the following expression
\bea
L_{\mu\nu} = 2 k_{1\mu} k_{2\nu} + 2 k_{1\nu} k_{2\mu} - Q^2 g_{\mu\nu},
\eea
while the hadronic tensor $W^{\mu\nu}$ can be written as
\bea
W^{\mu\nu}_h &= \int d^4 x \,e^{iq\cdot x} \langle 0|J^\mu(x) |hX\rangle\langle hX|J^\nu(0)|0\rangle,
\\
W^{\mu\nu}_{\rm jet} &= \int d^4 x \,e^{iq\cdot x} \langle 0|J^\mu(x) |JX\rangle\langle JX|J^\nu(0)|0\rangle,
\eea
where again a summation over the final-state unobserved particles $X$ is implied. 

\bef
\includegraphics[width=5.2in]{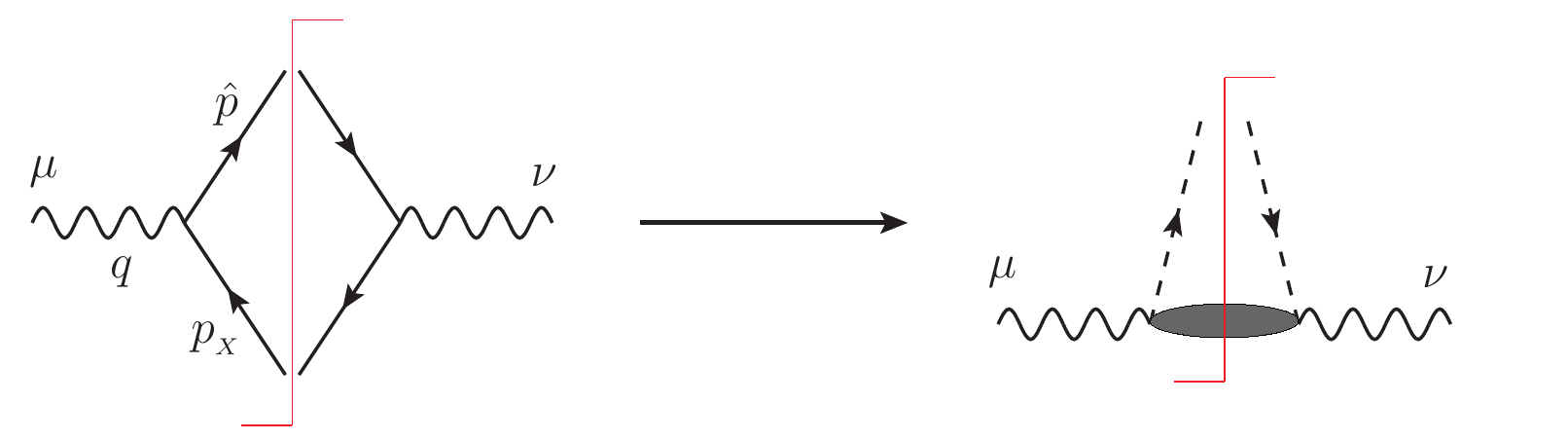}
\caption{Tree-level matching onto the operators for single inclusive hadron/jet production in $e^+e^-\to h X$ or $e^+e^-\to {\rm jet} X$. The red vertical line is the final-state cut.}
\label{fig:match}
\eef

In the region of phase space under consideration $p_X^2\sim Q^2$, the hard fluctuations $\sim p_X^2$ can be integrated out. Operationally this means we match $W_{\mu\nu}$ onto local operators in the effective theory which involves only the collinear fields in the direction of the hadron/jet, as illustrated in Fig.~\ref{fig:match}. This technique is the same as the one that has been used in~\cite{Bauer:2002nz}, for inclusive deep inelastic scattering in the so-called operator product expansion (OPE) region, inclusive Drell-Yan production, or heavy quark production in~\cite{Fickinger:2016rfd}. Following this seminal work, we have
\bea
W_h^{\mu\nu} \to & \int d\omega d\omega' \bigg[H_{e^+e^-\to q}^{\mu\nu}(\omega, \omega') 
{\rm Tr}\left(\frac{\sla{\bar n}}{2}\langle 0|\chi_{n,\omega}|hX\rangle\langle hX| \bar\chi_{n,\omega'}|0\rangle \right)
\nnu
&- H_{e^+e^-\to g}^{\mu\nu}(\omega, \omega')  \langle 0| {\mathcal B}_{n\perp, \omega}^{\mu}|hX\rangle\langle hX| {\mathcal B}_{n\perp, \omega'\, \mu} |0\rangle \bigg] + {\mathcal O}\left(m_h^2/p_T^2\right), 
\\
W_{\rm jet}^{\mu\nu} \to & \int d\omega d\omega' \bigg[H_{e^+e^-\to q}^{\mu\nu}(\omega, \omega') 
{\rm Tr}\left(\frac{\sla{\bar n}}{2} \langle 0|\chi_{n,\omega}|JX\rangle\langle JX| \bar\chi_{n,\omega'}|0\rangle \right)
\nnu
&- H_{e^+e^-\to g}^{\mu\nu}(\omega, \omega')  \langle 0| {\mathcal B}_{n\perp, \omega}^{\mu}|JX\rangle\langle JX| {\mathcal B}_{n\perp, \omega'\, \mu} |0\rangle \bigg] + {\mathcal O}\left(\mu_J^2/p_T^2\right).
\eea
The above factorization is simply a separation of physics at two different scales. For the hadron case, it is the scale of hadronization, i.e. the hadron mass $m_h$ and the scale of hard collisions $\sim p_T$. For  jet production, it is the natural scale of the jet function $\mu_J\sim p_{T R}$ and the scale of the hard collisions $p_T$. As $\mu_J^2/p_T^2 \approx R^2$, our factorization is valid up to the power corrections of jet radius $R$. 

To proceed further, one realizes that~\cite{Bauer:2002nz,Fickinger:2016rfd}, for single hadron production
\bea
\frac{1}{2N_c} {\rm Tr}\left(\frac{\sla{\bar n}}{2}\langle 0|\chi_{n,\omega}|hX\rangle\langle hX| \bar\chi_{n,\omega'}|0\rangle \right) &=  \int_0^1 \frac{dz_c}{z_c} \delta(\omega_{-}) \delta\left(z_c - \frac{2\bar n\cdot p_h}{\omega_{+}}\right) D_{q}^h(z_c),
\\
\frac{1}{2(N_c^2-1)}\langle 0| {\mathcal B}_{n\perp, \omega}^{\mu}|hX\rangle\langle hX| {\mathcal B}_{n\perp, \omega'\, \mu} |0\rangle &= -\frac{2}{\omega_{+}} \int_0^1 \frac{dz_c}{z_c} \delta(\omega_{-}) \delta\left(z_c - \frac{2\bar n\cdot p_h}{\omega_{+}}\right) D_{g}^h(z_c),
\eea
and for single jet production,
\bea
\frac{1}{2N_c} {\rm Tr}\left(\frac{\sla{\bar n}}{2}\langle 0|\chi_{n,\omega}|JX\rangle\langle JX| \bar\chi_{n,\omega'}|0\rangle \right) &=  \int_0^1 \frac{dz_c}{z_c} \delta(\omega_{-}) \delta\left(z_c - \frac{2\bar n\cdot p_J}{\omega_{+}}\right) J_{q}(z_c, \omega_J),
\\
\frac{1}{2(N_c^2-1)}\langle 0| {\mathcal B}_{n\perp, \omega}^{\mu}|JX\rangle\langle JX| {\mathcal B}_{n\perp, \omega'\, \mu} |0\rangle &= -\frac{2}{\omega_{+}} \int_0^1 \frac{dz_c}{z_c} \delta(\omega_{-}) \delta\left(z_c - \frac{2\bar n\cdot p_J}{\omega_{+}}\right) J_{g}(z_c, \omega_J),
\eea
where $\omega_{\pm} = \omega \pm \omega'$. After substituting the above expressions back into Eqs.~\eqref{eq:h1} and \eqref{eq:j1}, i.e., contracted with the leptonic tensor, we end up with the factorized forms as given in Eqs.~\eqref{eq:hadron} and \eqref{eq:jet}, for single hadron and single jet production, respectively. In other words, the short-distance hard functions $H_{e^+e^-\to c}$ are simply given by the contraction of the leptonic tensor with the hadronic ones,
\bea
H_{e^+e^-\to c}^h &\propto L_{\mu\nu} H_{e^+e^-\to c}^{\mu\nu, \,h} (\omega_{+} = 2 \bar n\cdot p_h/z_c, \omega_{-}= 0),
\\
H_{e^+e^-\to c}^{\rm jet} &\propto L_{\mu\nu} H_{e^+e^-\to c}^{\mu\nu, \,\rm jet} (\omega_{+} = 2 \bar n\cdot p_J/z_c, \omega_{-} = 0).
\eea

\subsection{NLO calculations: single hadron}
We will now compute in perturbation theory the short-distance hard functions $H_{e^+e^-\to c}^{h}$ and $H_{e^+e^-\to c}^{\rm jet}$, and will demonstrate that they are the same to  NLO accuracy. This is a standard matching calculation, where one replaces the hadron or the jet by a parton state on both sides, and one calculates both sides in an expansion of the strong coupling constant $\alpha_s$. For single inclusive hadron production, the NLO results are well-known~\cite{Altarelli:1979kv,Furmanski:1981cw,Nason:1993xx,Kretzer:2000yf,Anderle:2012rq}. It is convenient to write 
\bea
H_{e^+e^-\to c}^{h}(\hat p, \mu) = \frac{2\hat p_{T}}{s} \frac{d\hat\sigma_c(s,\hat p_T,\eta,\mu)}{dvdz}, 
\eea
and thus the cross section for $e^+e^-\to hX$ can be expressed as
\bea
\frac{d\sigma^{e^+e^-\to hX}}{dp_Td\eta}=\frac{2 p_T}{s}\sum_{c=q,\bar q,g}\int^1_{z_c^{\rm min}} \frac{dz_c}{z_c^2}\frac{d\hat\sigma_c(s,\hat p_T,\eta,\mu)}{dvdz}D^h_c(z_c,\mu),
\eea
where $\hat p_T = p_T/z_c$ and $z^{\rm min}_c= 2p_T\cosh \eta/\sqrt{s}$. At the same time we define the $v$ and $z$ variables as
\bea
v=1-\f{2\hat p_T}{\sqrt{\hat s}}e^{-\hat\eta}, \qquad z=\f{2\hat p_T}{\sqrt{s}}\cosh\eta \, .
\eea
Now the partonic cross section up to the NLO can be written as
\bea
\frac{d\hat\sigma_c}{dvdz} = \frac{d\hat\sigma_c^{(0)}}{dv}\delta(1-z)+\frac{\alpha_s(\mu)}{2\pi} \frac{d\hat\sigma_c^{(1)}}{dvdz},
\label{eq:lo+nlo}
\eea
where we have the leading order result 
\bea
\frac{d\hat\sigma_c^{(0)}}{dv} = \frac{N_c\, e_q^2\pi \alpha^2}{s} 2(v^2+(1-v)^2),
\eea
and the NLO expressions for both quark and gluon channels within the $\overline{\rm MS}$ scheme are given by, 
\bea
\frac{d\hat\sigma_q^{(1)}}{dvdz}  = & \frac{N_c\, e_q^2\pi \alpha^2}{s} C_F \left[2(v^2+(1-v)^2) \left((1+z^2)\left(\frac{\ln(1-z)}{1-z}\right)_+ - \frac{P_{qq}(z)}{C_F}\left(\ln\left(\frac{\mu^2}{s}\right)+\frac{3}{4}\right)\right. \right. 
\nnu
& \left. + \left(\frac{2\pi^2}{3}-\frac{27}{8}\right)\delta(1-z) + 2 \frac{1+z^2}{1-z}\ln z \right) + 2 \frac{1+z^2}{z^3} \left(\ln(1-z)+2\ln z-\ln\left(\frac{\mu^2}{s}\right) \right) 
\nnu
& \times \left(2 v^2 (z^2+z+1)-2 v (z^2+z+2)+z^2+2 \right) -\frac{1}{2z^3} \left(2 v^2 (3 z^4+3 z^3+6 z^2 \right. 
\nnu
& \left. +12 z+8) - 2 v (3 z^4+3 z^3+16 z+16)+3 z^4+9 z^3-12 z^2+8 z+16\right) \Bigg],
\\
\frac{d\hat\sigma_g^{(1)}}{dvdz}  = & \frac{N_c\, e_q^2\pi \alpha^2}{s} C_F \left[ 4\frac{z^2-2 z+2}{z^4} (2 v^2 + 2 v z - 4 v + z^2 - 2 z + 2) \right. 
\nnu 
& \left. \times\left(\ln(1-z)+2\ln z -\ln\left(\frac{\mu^2}{s}\right) \right) - 8 \frac{1-z}{z^4}  (6 v^2 + 6 v z - 12 v + z^2 - 6 z + 6)  \right].
\eea

\subsection{NLO calculation: single jet}
Let us now turn to the calculations of the short-distance hard functions for single inclusive jet production. At LO, a single parton makes the jet and the semi-inclusive jet functions are given by $J_i(z, \omega_J, \mu) = \delta(1-z)$. The short-distance hard functions are calculated from the standard $e^+e^-\to q\bar q$ channel, and they are the same for single hadron and jet production, we thus obtain at LO
\bea
d\hat \sigma_{c}^{(0),\, \rm jet} = d\hat \sigma_{c}^{(0), h}\equiv d\hat \sigma_{c}^{(0)},
\eea
which is given in Eq.~\eqref{eq:lo+nlo}. At NLO, the calculations are more involved. To produce analytical calculations, we will use the narrow jet approximation, which is equivalent to requiring the jet to be highly collimated, as it is usually assumed in the SCET computations. We will follow the computations in~\cite{Mukherjee:2012uz,Jager:2004jh}, where one starts from the NLO single-parton inclusive cross section (i.e. $d\hat \sigma_{e^+e^-\to cX}$ for $e^+e^-$ collisions), relevant for the single-inclusive hadron production, $e^+e^-\to hX$, as calculated above, and convert these results to the desired single-inclusive jet cross sections. The procedure is straightforward, and has been explained in detail in~\cite{Mukherjee:2012uz,Jager:2004jh}. Here we recall these results for completeness and for later convenience when we perform the matching onto the semi-inclusive jet functions to obtain the short-distance hard functions. 

\bef
\includegraphics[width=4.8in]{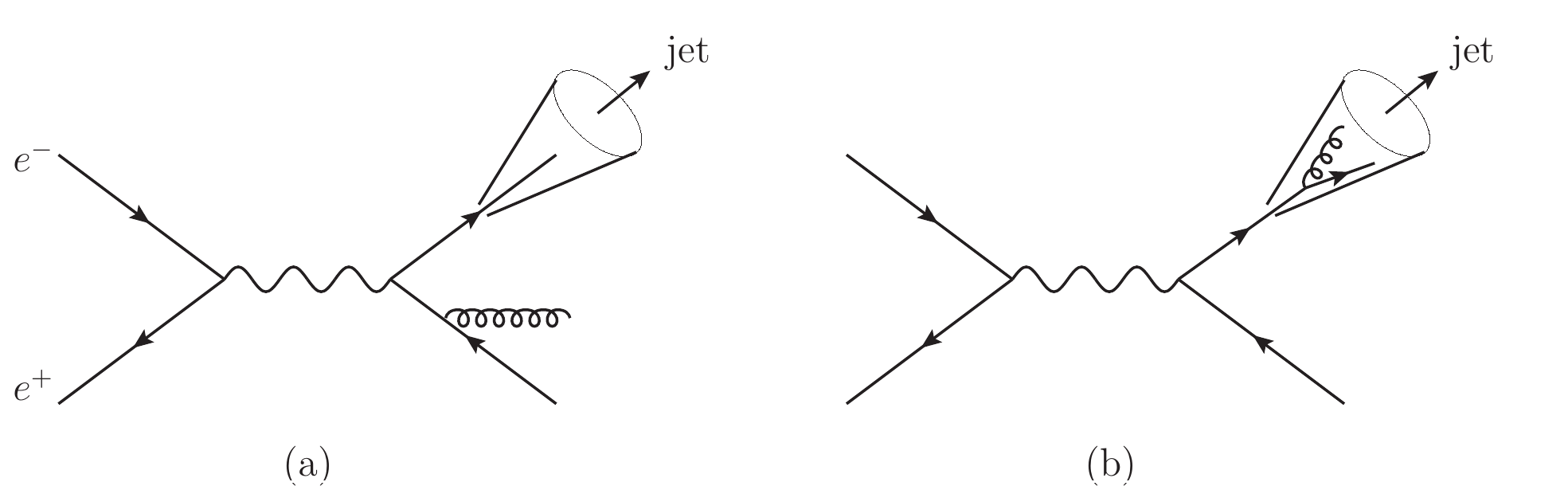}
\caption{Contributions to the single-inclusive jet cross section from partonic scattering: (a) with only one parton inside the jet, (b) two essentially collinear partons, $q$ and $g$, form a narrow jet.}
\label{fig:jet-config}
\eef

In order to convert analytically the single-parton inclusive cross sections to single inclusive jet cross sections, we use the narrow jet approximation and the fact that the jet is formed either by a single final-state parton or jointly by two partons, as illustrated in Fig.~\ref{fig:jet-config}. The final expression for the desired partonic jet cross section can be written as~\footnote{Note we do not have the situation where $q$ and $\bar q$ forms the jet together at leading power~\cite{Kang:2011mg,Kang:2014tta,Kang:2014pya}, since gluons do not interact directly with electrons/photons.}
\bea
d\hat \sigma_{e^+e^-\to {\rm jet} X} =& \big[d\hat \sigma_q - d\hat \sigma_{q(g)}\big]  +   \big[d\hat \sigma_g - d\hat \sigma_{g(q)}\big] + d\hat \sigma_{qg}
+ \left(q\to \bar q \right),
\label{eq:formjet}
\eea
where we suppressed a term for anti-quark $\bar q$, and $d\hat \sigma_{q}$ is the single quark inclusive cross section as given above, while $d\hat \sigma_{q(g)}$ is the cross section where still $q$ is observed, but $g$ is also in the cone. Thus their difference $d\hat \sigma_q - d\hat \sigma_{q(g)}$ gives exactly the configuration where only $q$ forms the jet, while $g$ is outside the jet cone. Similarly for $d\hat \sigma_g - d\hat \sigma_{g(q)}$ when only $g$ forms the jet while $q$ is outside the jet cone. On the other hand, $d\hat \sigma_{qg}$ is the cross section where $q$ and $g$ are both inside the cone and form the jet together. In other words, Eq.~\eqref{eq:formjet} produces exactly the contributions as illustrated in Fig.~\ref{fig:jet-config}. 

It may be important to emphasize that the single-parton inclusive cross sections $d\hat \sigma_{q}$ and $d\hat \sigma_{g}$ are obtained after a subtraction of final-state collinear singularities in the $\overline{\rm MS}$ scheme. Thus upon calculation of the combinations $-d\hat \sigma_{q(g)}-d\hat \sigma_{g(q)} + d\hat \sigma_{qg}$ in the above equation, one also needs to perform an $\overline{\rm MS}$ subtraction to compensate the aforementioned subtraction and thus obtain the correct combination, for details, see~\cite{Mukherjee:2012uz,Jager:2004jh}. The way to compute $d\hat \sigma_{q(g)}$ and $d\hat \sigma_{g(q)}$ are given in~\cite{Jager:2004jh}. Since there is only one parton inside the jet, there is no jet algorithm dependence. On the other hand, the cross section $d\hat \sigma_{qg}$ represents the situation where both partons $q$ and $g$ jointly form the jet, and it will depend on the jet algorithm. All of them $d\hat \sigma_{q(g)}$, $d\hat \sigma_{g(q)}$, and $d\hat \sigma_{qg}$ are proportional to the lowest order cross section, with the detailed expressions given in~\cite{Jager:2004jh,Mukherjee:2012uz} for both cone and \kt~jets. We find that they can be cast in the following form:  
\bea
-d\hat \sigma_{q(g)} &= d\hat\sigma_q^{(0)} \otimes J_{q\to q(g)}(z_c, \omega_J),
\\
-d\hat \sigma_{g(q)} &= d\hat\sigma_q^{(0)} \otimes J_{q\to (q)g}(z_c, \omega_J),
\\
d\hat \sigma_{qg} &= d\hat\sigma_q^{(0)} \otimes J_{q\to qg}(z_c, \omega_J),
\eea
where $\omega_J = 2p_T\cosh \eta$ is the jet energy and $\otimes$ represents the standard convolution over the momentum fraction $z_c$. We, thus, obtain 
\bea
-d\hat \sigma_{q(g)}  -d\hat \sigma_{g(q)} + d\hat \sigma_{qg}  &= d\hat\sigma_q^{(0)} 
\otimes \big[J_{q\to q(g)}(z_c, \omega_J) + J_{q\to (q)g}(z_c, \omega_J) + J_{q\to qg}(z_c, \omega_J)\big],
\nnu
&= d\hat\sigma_q^{(0)}\otimes J_q^{(1)}(z_c, \omega_J).
\eea
In the second step, we have used Eq.~\eqref{eq:J1qterms}. At the same time, with an additional $\overline{\rm MS}$ subtraction as discussed above to compensate the same subtraction performed for $d\hat \sigma_{q}$, we have
\bea
\big[-d\hat \sigma_{q(g)}  -d\hat \sigma_{g(q)} + d\hat \sigma_{qg}\big]_{\overline{\rm MS}}  = d\hat\sigma_q^{(0)}\otimes J_q^{(1)}(z_c, \omega_J, \mu),
\eea
where $J_q^{(1)}(z_c, \omega_J, \mu)$ is the renormalized quark jet function given in Eq.~\eqref{eq:ren-Jq}. 

Finally, realizing that the single-parton inclusive cross section can be written as a trivial convolution with a $\delta(1-z)$ function, we can write
\bea
d\hat \sigma_c = d\hat\sigma_c^{(1)} \otimes J_c^{(0)}(z_c, \omega_J, \mu), 
\eea
with $ J_c^{(0)}(z_c, \omega_J, \mu) = \delta(1-z_c)$. We can then rewrite Eq.~\eqref{eq:formjet} up to the NLO as follows
\bea
d \sigma_{e^+e^-\to {\rm jet} X} =& \big[d\hat\sigma_q^{(0)} + d\hat\sigma_q^{(1)}\big] \otimes J_q^{(0)}(z_c, \omega_J, \mu) + d\hat\sigma_q^{(0)}\otimes J_q^{(1)}(z_c, \omega_J, \mu) 
\nnu
&+  d\hat\sigma_g^{(1)} \otimes J_g^{(0)}(z_c, \omega_J, \mu) + (q\to \bar q).
\eea
This is exactly the perturbative expansion up to NLO of our factorized formula given in Eq.~\eqref{eq:jet}, i.e.
\bea
d \sigma_{e^+e^-\to {\rm jet} X} =& \sum_c d\hat\sigma_c \otimes J_c(z_c, \omega_J, \mu)
\label{eq:jet-final}
\\
=& \sum_c \big[d\hat\sigma_c^{(0)} + d\hat\sigma_c^{(1)}\big] \otimes [J_c^{(0)}(z_c, \omega_J, \mu)+ J_c^{(1)}(z_c, \omega_J, \mu)],
\eea
where we drop ${\mathcal O}(\as^2)$ contributions that appear in the form of $d\hat\sigma_c^{(1)}\otimes J_c^{(1)}$ above. Eq.~\eqref{eq:jet-final} clearly demonstrates that the short-distance hard functions are exactly the same as those for single hadron production up to NLO. Even though we did not perform the matching calculations beyond NLO, and thus cannot make a definite statement 
but we conjecture that such a conclusion remains true even beyond the NLO. This is because the short-distance hard functions only depend on the hard scale $\mu\sim p_T$ (not on the lower scale associated with jet $\mu_J\sim p_{T R}$). Within $\overline{\rm MS}$ scheme, there seems no other way around. Of course this could be checked through explicit calculations. 

\section{Phenomenology: $pp\to {\rm jet} X$}
\label{sec:pp}
In this section, we show phenomenological applications for single inclusive jet production in $pp$ collisions at the LHC. In particular, we present how the resummation of logarithms of the small jet radius affects the inclusive jet cross sections. 

\subsection{Matching NLO and $\ln R$ resummation}
Following our discussion on the factorization formalism for $e^+e^-\to {\rm jet} X$, we can easily generalize the formula to write the cross section for $pp\to {\rm jet}X$ as
\bea
\label{eq:sigjetX}
\frac{d\sigma^{pp\to {\rm jet}X}}{dp_Td\eta}  = & \frac{2 p_T}{s}\sum_{a,b,c}\int_{x_a^{\rm min}}^1\f{dx_a}{x_a}f_a(x_a,\mu)\int_{x_b^{\rm min}}^1\f{dx_b}{x_b} f_b(x_b,\mu) 
\nnu
&\times \int^1_{z_c^{\rm min}} \frac{dz_c}{z_c^2}\frac{d\hat\sigma^c_{ab}(\hat s,\hat p_T,\hat \eta,\mu)}{dvdz}J_c(z_c,\omega_J,\mu).
\eea
Such a factorized formula has already been conjectured in~\cite{Kaufmann:2015hma}, if one chooses the fixed NLO results for $J_c(z_c,\omega_J,\mu)$ as given in Eqs.~\eqref{eq:ren-Jq} and \eqref{eq:ren-Jg}. Here $s$, $p_T$ and $\eta$ correspond to the center of mass (CM) energy, the jet transverse momentum and jet rapidity, respectively. The hard functions $d\hat\sigma_{ab}^c(\hat s,\hat p_T,\hat\eta,\mu)$ are functions of the corresponding partonic variables: the partonic CM energy $\hat s=x_ax_bs$, the partonic transverse momentum $\hat p_T=p_T/z_c$ and the partonic rapidity $\hat\eta=\eta-\ln(x_a/x_b)/2$. The variables $v,z$ can be expressed in terms of these partonic variables
\bea
v=1-\f{2\hat p_T}{\sqrt{\hat s}}e^{-\hat\eta}, \qquad z=\f{2\hat p_T}{\sqrt{s}}\cosh\hat\eta \, .
\label{eq:vz}
\eea
Up to one loop, the hard functions take the form
\bea
\frac{d\hat\sigma_{ab}^c}{dvdz} = \frac{d\hat\sigma_{ab}^{c,(0)}}{dv}\delta(1-z)+\frac{\alpha_s(\mu)}{2\pi} \frac{d\hat\sigma_{ab}^{c,(1)}}{dvdz}.
\eea
As demonstrated above, the hard functions here are the same as the hard functions for the process $pp\to hX$. The corresponding expressions were presented in~\cite{Aversa:1988vb,Jager:2002xm}. Finally, the integration limits in~(\ref{eq:sigjetX}) are customarily written in terms of the hadronic variables $V,Z$,
\be
V=1-\f{2p_T}{\sqrt{s}}e^{-\eta}, \qquad Z=\f{2 p_T}{s}\cosh\eta \, ,
\ee
and are given by
\be
x_a^{\rm min}=1-\f{1-Z}{V},\quad x_b^{\rm min}=\f{1-V}{1+(1-V-Z)/x_a},\quad z_c^{\rm min}=\f{1-V}{x_b}-\f{1-V-Z}{x_a} \, .
\ee

With our evolution equations for the semi-inclusive jet functions, $J_{q,g}(z,\omega_J,\mu)$, which can be evolved from scale $\mu_J = p_{T R}$ to the scale $\mu \sim p_T$ as in Eq.~\eqref{eq:DGLAP3}, we can resum the large logarithms of the jet radius $\ln R$. For phenomenological predictions, it is also necessary to combine the $\ln R$ resummation with the results from the fixed-order calculations. For  concreteness, in most of the  discussion in the rest of the paper we will perform DGLAP evolution for the semi-inclusive jet functions with LO ${\mathcal O}(\alpha_s)$ splitting functions as given in Sec.~\ref{sec:solution}, commonly referred as leading logarithmic resummation (LL$_{R}$). At the end of the section, we comment on next-to-leading logarithmic resummation (NLL$_{R}$). In order to combine NLO and LL$_{R}$ results, we write the inclusive jet cross section in Eq.~\eqref{eq:sigjetX} 
schematically as
\bea
\label{eq:NLOmatching}
d\sigma^{pp\to {\rm jet}X} \sim & \left(d\hat\sigma^{c,(0)}_{ab} + d\hat\sigma^{c,(1)}_{ab} \right)\otimes \left(J_c^{(0)}+J_c^{(1)} \right) 
\nnu
=& \left(d\hat\sigma^{c,(0)}_{ab} + d\hat\sigma^{c,(1)}_{ab} \right) \otimes J_c^{(0)} + d\hat\sigma^{c,(0)}_{ab} \otimes J_c^{(1)} + {\cal O}(\as^2)\, ,
\eea
where the term $d\hat\sigma^{c,(1)}_{ab}\otimes J_c^{(1)}$ is at ${\cal O}(\as^2)$, i.e., part of NNLO contribution, and will be dropped for consistency. This allows us to get back to the NLO calculation of~\cite{Mukherjee:2012uz} in the limit of having no evolution for the semi-inclusive jet function. At the same time, when we evolve both $J_c^{(0)}$ and $J_c^{(1)}$ through our DGLAP evolution equations Eq.~\eqref{eq:DGLAP3} from $\mu_J = p_{T R}$ to $\mu \sim p_T$, we are resumming the logs of $R$. Since the initial scale of the evolution depends on $R$, we obtain the limit of no evolution for $R\to 1$. Even though the limit of no evolution, $R\to 1$, is beyond the approximation of narrow jets, it serves as an important numerical cross check of our DGLAP-based resummation code.

\subsection{Dealing with the semi-inclusive jet function at $z\to 1$}
As can be seen already from Fig.~\ref{fig:Jqevo}, the evolved semi-inclusive jet functions are still divergent for $z\to 1$. Therefore, we can not directly use them in order to calculate a cross section. For example, for $pp\to{\rm jet}X$, we would have to integrate the jet functions over $z_c$ up to one, where they are divergent. We would like to emphasize again that the evolution does not render the initially divergent distributions finite for $z=1$. We deal with this issue by adopting a prescription developed in the context of fragmentation functions for quarkonia in~\cite{Bodwin:2015iua}. The main idea is to separate the integral in Eq.~\eqref{eq:sigjetX} into two pieces by introducing a cutoff $\varepsilon$. This way, we can integrate part of the cross section analytically instead of numerically. Schematically, we have
\be\label{eq:separate}
\int^1_{z_c^{\rm min}} \frac{dz_c}{z_c^2}\frac{d\hat\sigma_{ab}^c(z_c)}{dvdz}J_c(z_c) = \int^{1-\varepsilon}_{z_c^{\rm min}} \frac{dz_c}{z_c^2}\frac{d\hat\sigma_{ab}^c(z_c)}{dvdz}J_c(z_c) + \int^1_{1-\varepsilon} \frac{dz_c}{z_c^2}\frac{d\hat\sigma_{ab}^c(z_c)}{dvdz}J_c(z_c) \,,
\ee
where we have left the dependence on other variables than $z_c$ implicit to shorten our notation. Note that the variables $v$ and $z$ depend on $z_c$ as specified in Eq.~\eqref{eq:vz}. The cutoff parameter $\varepsilon$ is a small positive number chosen such that the the first integral can be computed numerically up to $1-\varepsilon$ using the evolved semi-inclusive jet functions. Our final numerical results are in fact independent of the choice of $\varepsilon$ to a remarkable degree. On the other hand, following~\cite{Bodwin:2015iua}, we rewrite the second term in Eq.~(\ref{eq:separate}) as
\bea
&\int^1_{1-\varepsilon} \frac{dz_c}{z_c^2}\frac{d\hat\sigma_{ab}^c(z_c)}{dvdz}J_c(z_c) = \int^1_{1-\varepsilon} \frac{dz_c}{z_c^2}\left[\frac{d\hat\sigma_{ab}^c(z_c)}{dvdz}\, z_c^{-N}\right] \left[z_c^N J_c(z_c)\right] \nnu
&\approx  \left[\frac{d\hat\sigma_{ab}^c(z_c)}{dvdz} \right]_{z_c=1} \times \int^1_{1-\varepsilon} dz_c\, z_c^{N-2} \,J_c(z_c) \nnu
 & =  \left[\frac{d\hat\sigma_{ab}^c(z_c)}{dvdz} \right]_{z_c=1} \times \left[\int^1_0 dz_c\, z_c^{N-2} \,J_c(z_c) - \int^{1-\varepsilon}_0 dz_c\, z_c^{N-2} \,J_c(z_c)  \right].
\label{eq:N-term}
\eea
Here, we purposely multiply the semi-inclusive jet function $J_c(z_c)$ by a factor $z_c^N$ to ensure that the second factor in the second line is finite over the integration region, which is true as long as $N>2$. The approximation in the second line is obtained by expanding $z_c^{-N}d\hat\sigma_{ab}^{c}(z_c)/dvdz$ in powers of $1-z_c$ and keeping only the first term in the expansion. In the last line, the first term in the bracket can be calculated numerically and it is simply given by the $N-1$ Mellin moments of the evolved semi-inclusive jet function. In practice, we can obtain this part from our evolution code before the Mellin inverse is taken. On the other hand, the second term in the bracket is given by the truncated $N-1$ Mellin moments of the evolved semi-inclusive jet functions, which can be calculated numerically as it only requires the $J_c(z_c)$ for $z_c<1-\varepsilon$ as input. For this approach to work,  Eq.~\eqref{eq:N-term}  should be independent of the choice of $N$. We find that the numerical results change only $\sim 0.01\%$ for $N$ in the range of $N=3-7$~\cite{Bodwin:2015iua,Dasgupta:2014yra}. To summarize, we calculate the single inclusive jet cross section in the following way
\bea
& \int^1_{z_c^{\rm min}} \frac{dz_c}{z_c^2}\frac{d\hat\sigma_{ab}^c(z_c)}{dvdz}J_c(z_c) \approx \int^{1-\varepsilon}_{z_c^{\rm min}} \frac{dz_c}{z_c^2}\frac{d\hat\sigma_{ab}^c(z_c)}{dvdz}J_c(z_c) \nnu
& + \left[\frac{d\hat\sigma_{ab}^c(z_c)}{dvdz} \right]_{z_c=1} \times \left[\int^1_0 dz_c\, z_c^{N-2} \,J_c(z_c) - \int^{1-\varepsilon}_0 dz_c\, z_c^{N-2} \,J_c(z_c)  \right].
\eea
We can test this prescription numerically by considering the case of almost no evolution, i.e. by choosing $R\to 1$ and then comparing with the calculations from a standard NLO code for jet cross sections~\cite{Aversa:1988vb,Jager:2004jh,Mukherjee:2012uz}. 

\subsection{Numerical results for the LHC}
\bef
\includegraphics[width=4.2in]{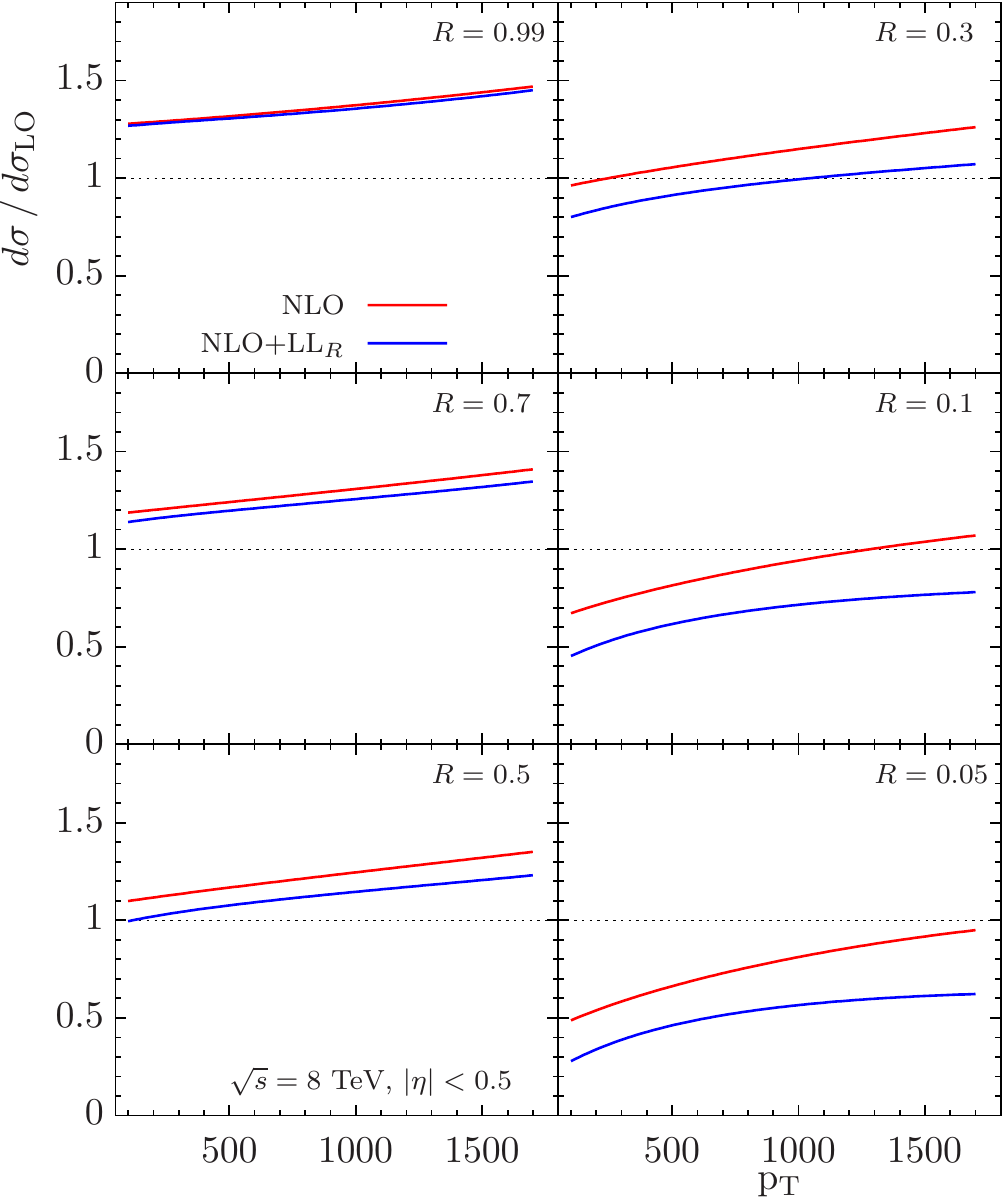}
\caption{NLO (red) and ${\rm NLO+LL}_R$ (blue) cross sections normalized to the leading-order result for different values of $R=0.99-0.05$. The small-$R$ approximation is only valid up to $R\approx 0.7$. However, $R=0.99$ illustrates that the resummed result does converge to the NLO result for $R\to 1$. As an example, we choose $\sqrt{s}=8$~TeV and $|\eta|<0.5$.}
\label{fig:cross-res}
\eef
We now turn to the numerical results for inclusive jet cross sections at the LHC. As an example, we choose a CM energy of $\sqrt{s}=8$~TeV and the jet rapidity $|\eta|<0.5$. We perform the numerical calculations using the CTEQ6.6M NLO parton distribution functions~\cite{Nadolsky:2008zw}. In Fig.~\ref{fig:cross-res}, we plot both NLO (red) and ${\rm NLO+LL}_R$ (blue) cross sections as a function of the jet transverse momentum $p_T$ for different values of $R=0.99-0.05$. Both cross sections are normalized to the leading-order result for better visualization. In the calculations, we take the nominal scale choices: both the renormalization scale $\mu_R$ (associated with $\alpha_s$) and the factorization scale $\mu_F$ (associated with the parton distributions functions in the incoming protons) are equal to $p_T$ of the jet, $\mu_R = \mu_F = p_T$; the natural scale for semi-inclusive jet functions $\mu_J = p_{TR}$ as given in Eq.~\eqref{eq:pTR}, which is further evolved to scale $\mu = p_T$. The small-$R$ approximation is only valid up to $R\approx 0.7$, see the detailed discussion in \cite{Mukherjee:2012uz,Dasgupta:2016bnd}. However, $R=0.99$ illustrates that the resummed result does converge to the NLO result for $R\to 1$, as can be seen clearly in the top left panel. We also find that when compared to the NLO results, ${\rm NLO+LL}_R$ results lead to about $10-20\%$ reduction in the cross section for the intermediate $R=0.3-0.5$. As $R$ becomes even smaller, the reduction becomes more evident. 
\bef
\includegraphics[width=4.2in]{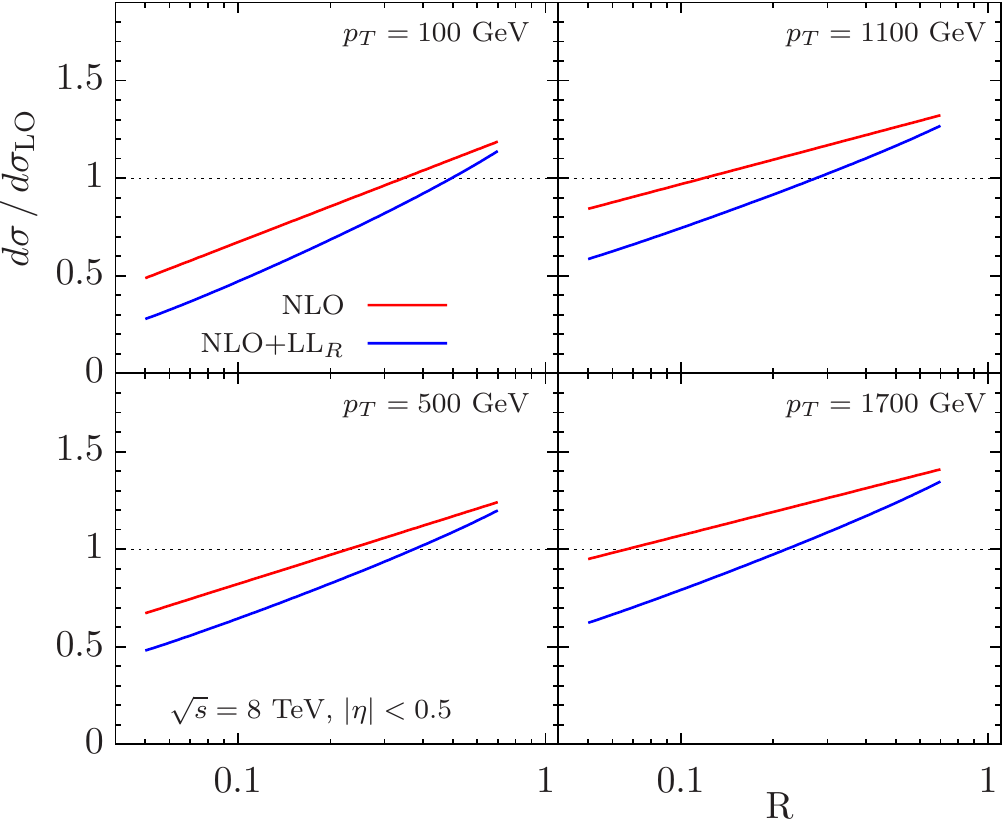}
\caption{The NLO (red) and ${\rm NLO+LL}_R$ (blue) cross sections normalized to the leading-order result are now shown as a function of $R$ for different values of the jet transverse momentum $p_T=100,\, 500,\, 1100,\, 1700$~GeV. Again, we choose $\sqrt{s}=8$~TeV and $|\eta|<0.5$. Note that here we chose to plot the ratio only until $R=0.7$ which is the uppermost value where the small-$R$ approximation is expected to be valid.}
\label{fig:cross-resR}
\eef

To see more clearly the reduction of the cross section as $R$ decreases, in Fig.~\ref{fig:cross-resR} we show the NLO (red) and ${\rm NLO+LL}_R$ (blue) cross sections normalized to the leading-order result, now as a function of the jet radius $R$ for different values of the jet transverse momentum $p_T=100$, 500, 1100, 1700 GeV, respectively. Again, we choose $\sqrt{s}=8$~TeV and $|\eta|<0.5$. Note that here we chose to plot the ratio only until $R=0.7$ which is the uppermost value where the small-$R$ approximation is expected to be valid. The reduction from the NLO result can be as large as $30-40\%$ at $p_T=1700$ GeV for a very small $R\sim 0.05$. 
\bef
\includegraphics[width=2.8in]{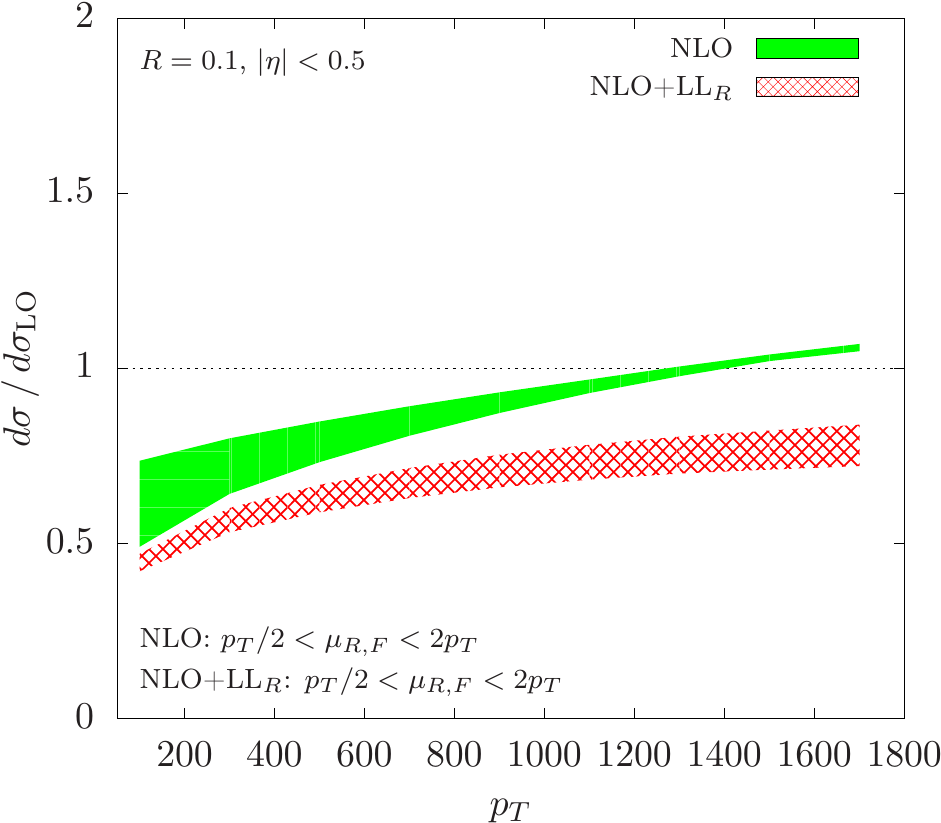} 
\hskip 0.2in
\includegraphics[width=2.8in]{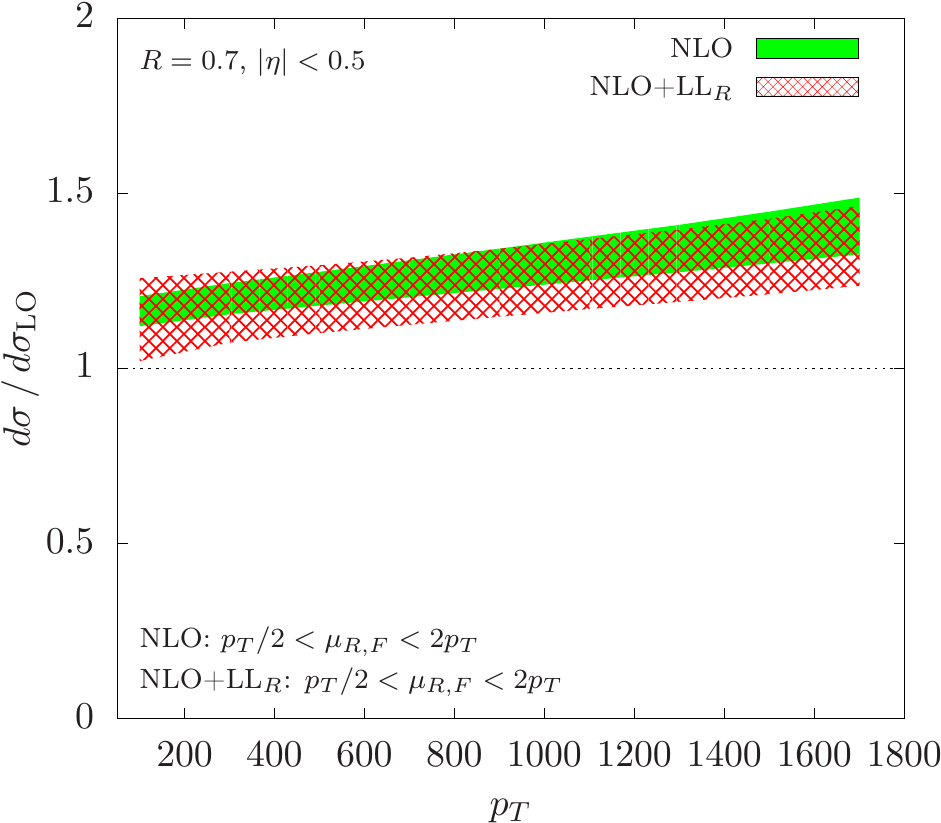} 
\caption{Comparison of the scale dependence of the NLO result (green) and the ${\rm NLO+LL}_R$ resummed calculation (red). Both calculations are normalized by the leading-order cross section. For a proper comparison, we vary in both cases only the renormalization and the factorization scales independently $p_T/2<\mu_{R,F}<p_T$ and take the envelope. Note that for the resummed calculation we keep the jet scale $\mu_J$ and the final scale of the DGLAP evolution fixed. We present results for $\sqrt{s}=8$~TeV, $|\eta|<0.5$ and $R=0.1$ (left), $R=0.7$ (right).}
\label{fig:NLO-LLR}
\eef

\bef
\includegraphics[width=2.8in]{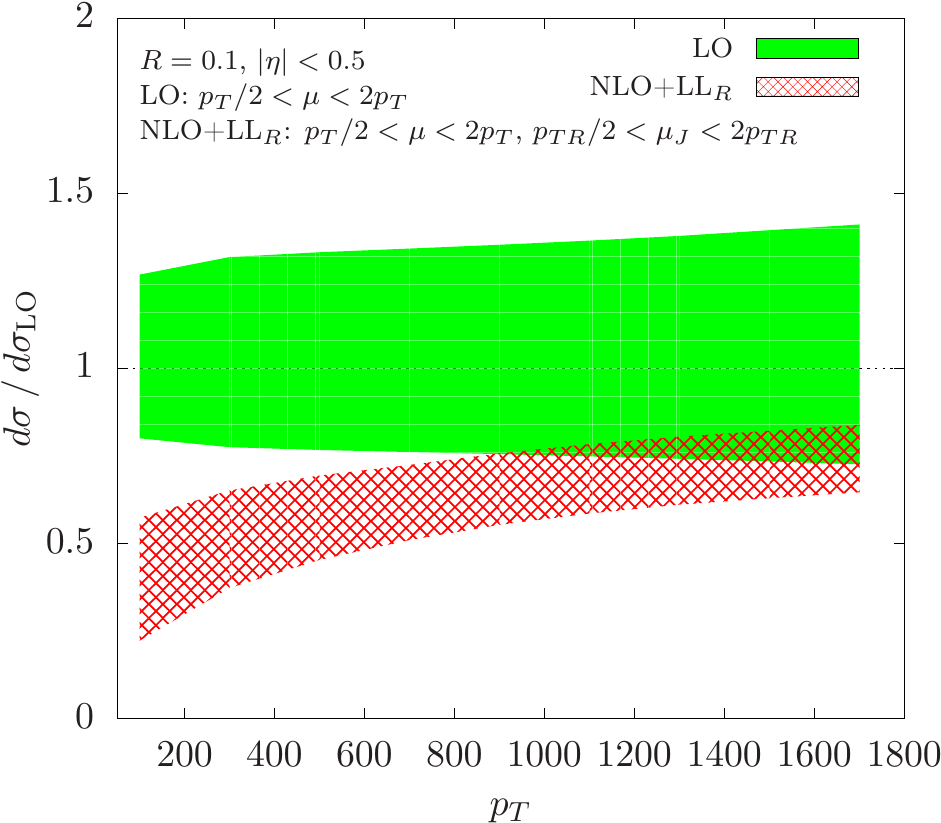} 
\hskip 0.2in
\includegraphics[width=2.8in]{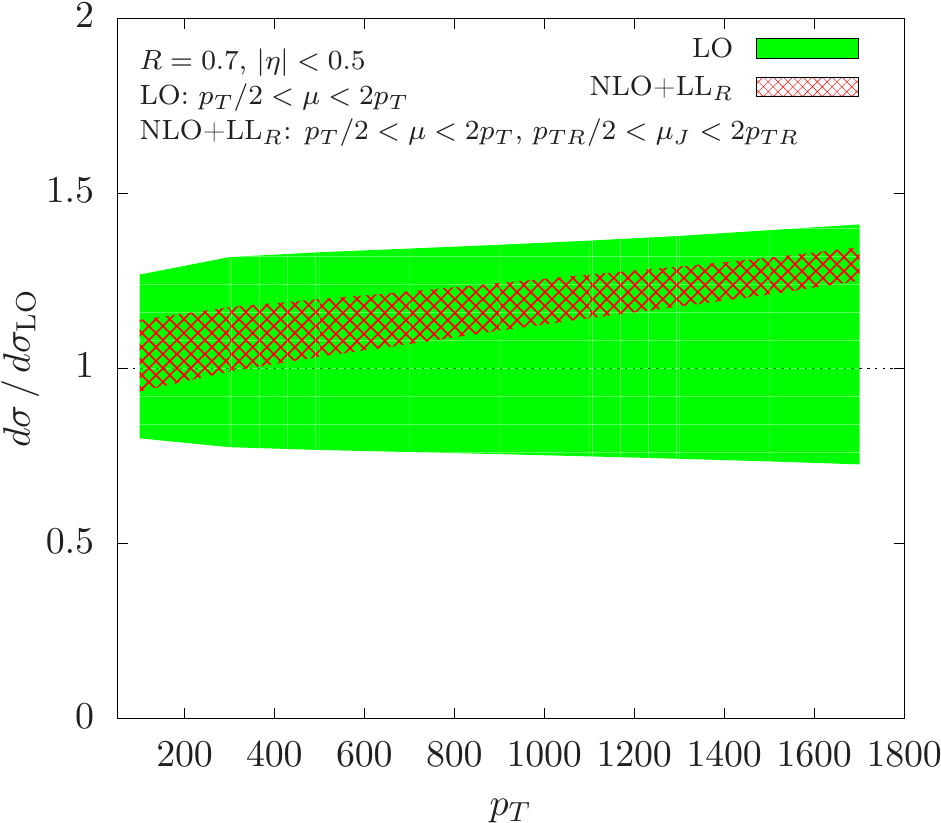} 
\caption{Comparison of the scale dependence of the LO result (green) and the ${\rm NLO+LL}_R$ resummed calculation (red). Both calculations are normalized by the nominal leading-order cross section. We vary the renormalization scale $p_T/2<\mu<2 p_T$, as well as the jet scale $p_{TR}/2<\mu_J< 2p_{TR}$ independently, and take the envelope. We present results for $\sqrt{s}=8$~TeV, $|\eta|<0.5$ and $R=0.1$ (left panel), $R=0.7$ (right panel).}
\label{fig:LO-NLO}
\eef

Let us now discuss the theoretical uncertainties of our factorization formalism, especially those from the sale variations. In Fig.~\ref{fig:NLO-LLR} we plot the scale uncertainty of the NLO result (green) and the ${\rm NLO+LL}_R$ resummed calculation (red). Both calculations are normalized by the LO cross section, with the LO result calculated at the nominal scales $\mu_{R} = \mu_{F} = p_T$. For  proper comparison, we vary in both cases only the renormalization and the factorization scales independently $p_T/2<\mu_{R,F}<p_T$ and take the envelope. Note that for the resummed calculation, we keep the jet scale $\mu_J$ and the final scale of the DGLAP evolution fixed at the nominal values: $\mu_J = p_{TR}$ and $\mu = p_T$. We present results for $\sqrt{s}=8$~TeV, $|\eta|<0.5$ and $R=0.1$ (left panel) and $R=0.7$ (right panel). As one can see, for the small jet radius $R=0.1$ case, there is a strong reduction in the cross section from the NLO+LL$_{\rm R}$ results in the high $p_T$ region, and the uncertainty bands for NLO and NLO+LL$_{R}$ results do not overlap. It might be worthwhile mentioning that the scale uncertainty of the NLO result for $R=0.1$ is extremely small in the high $p_T\gtrsim 1000$ GeV region. Such a small (almost vanishing) scale dependence is usually considered to be unphysical, likely to be an artifact of the NLO formalism, as advocated in \cite{Dasgupta:2016bnd}. However, such an unphysically small scale dependence does not appear in our $\ln R$-resummed NLO+LL$_{R}$ result, which has an uncertainty band of similar size in the whole $p_T$ region. 
\bef
\includegraphics[width=3.0in]{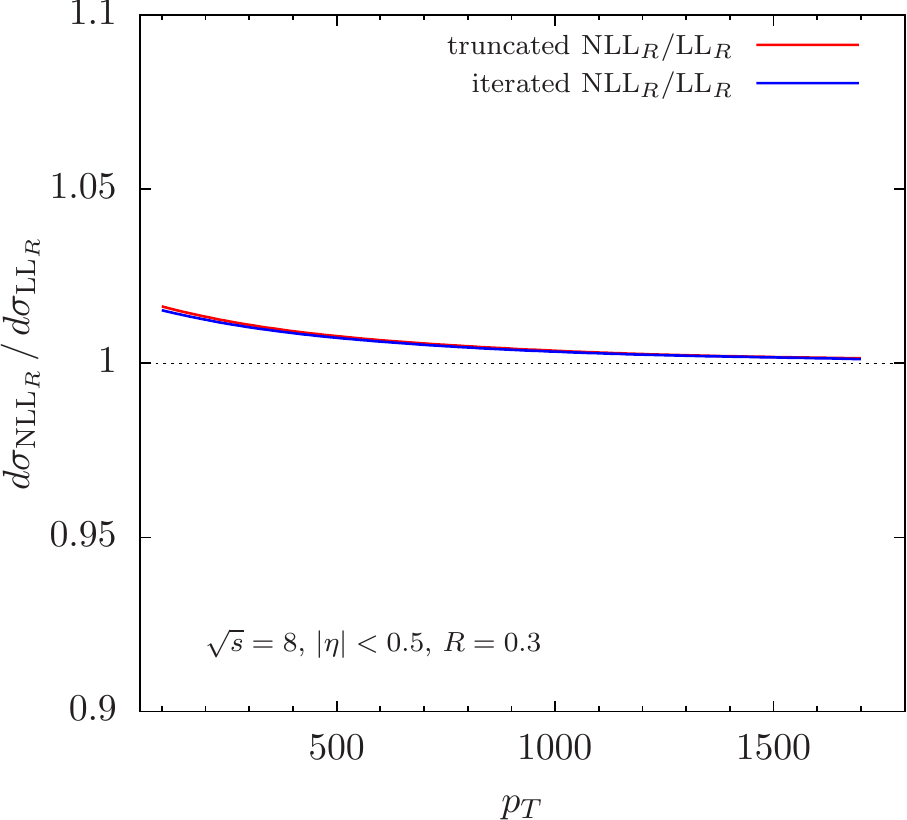}
\caption{The ratio of NLO+NLL$_R$ cross section over NLO+LL$_R$ result for jet radius $R=0.3$ is plotted as a function of jet transverse momentum $p_T$. There are two common solutions in Mellin moment space, and we plot both of them: truncated solution (red) and iterated solution (blue).}
\label{fig:LL-NLL}
\eef

Within SCET, the single inclusive jet cross section will eventually contain simply two scales. One is the renormalization scale $\mu$ for the hard function and the jet function~\footnote{One might simply consider this as the case when one chooses $\mu_R = \mu_F$ to be equal and varies them together.} as given in the factorization formalism in Eq.~\eqref{eq:sigjetX}. The other one is the scale $\mu_J$ that arises when we perform the $\ln R$ resummation, i.e., when we evolve the semi-inclusive jet function from the initial scale $\mu_J$ to the renormalization scale $\mu$. We vary both of them by a factor of 2 with respect to their natural values: $p_T/2 < \mu < 2p_T$ and $p_{TR}/2 < \mu_J < 2p_{TR}$. In Fig.~\ref{fig:LO-NLO}, we plot the scale uncertainty of the LO result (green) and the ${\rm NLO+LL}_R$ resummed calculation (red). Again both results are normalized to the LO cross section calculated at the nominal renormalization scale $\mu = p_T$. One clearly sees that the theoretical uncertainties are significantly reduced from the LO to the NLO+LL$_{R}$ results. 

So far we have presented NLO+LL$_{R}$ results. In fact we can also easily implement the NLO+NLL$_{R}$ cross sections. To do that, one starts from the matching formula in Eq.~\eqref{eq:NLOmatching}, and performs the NLO DGLAP evolution for the semi-inclusive jet functions $J_c^{(0,1)}$ by using NLO ${\mathcal O}(\alpha_s^2)$ splitting functions. One might recall that for a consistent NLO calculations of single hadron production, we usually use NLO-evolved fragmentation functions $D_i^h(z, \mu)$. In the same spirit, let us perform NLO-evolved semi-inclusive jet functions and assess their impact in the cross sections. In Fig.~\ref{fig:LL-NLL}, we plot the ratio of the NLO+NLL$_{R}$ result over NLO+LL$_{R}$ calculation for $R=0.3$ as a function of jet transverse momentum $p_T$. There are two common solutions in Mellin moment space~\footnote{For details, see Refs.~\cite{Vogt:2004ns,Anderle:2015lqa}.}, and we plot both of them: truncated solution (red) and iterated solution (blue). We find that such a ratio is only around $1\%$ level, indicating that the 
NLO+NLL$_R$ resummation does not provide significant effects on the inclusive jet cross sections compared with NLO+LL$_R$.  

\section{Summary}
\label{sec:summary}
In this paper, motivated by the need for small jet radius resummation for {\it inclusive} jet cross sections, we introduced a new kind of jet function: the semi-inclusive jet function $J_i(z, \omega_J, \mu)$. It describes the jet initiated by a parton $i$ which retains a momentum fraction $z$ of the parent parton energy. We demonstrated that it is these semi-inclusive jet functions for collinear quarks and gluons  that appear in the factorized formalism for the single inclusive jet cross sections. When implemented in the factorization formula, single inclusive jet production shares the same short-distance hard functions as single inclusive hadron production, with only the fragmentation functions $D_i^h(z, \mu)$ replaced by $J_i(z, \omega_J, \mu)$. Within Soft Collinear Effective Theory, we calculated both $J_q(z, \omega_J, \mu)$ and $J_g(z, \omega_J, \mu)$ to the next-to-leading order and demonstrated that the renormalization group equations of $J_i(z, \omega_J, \mu)$ follow exactly the usual timelike DGLAP evolution. Such RG equations can be used to perform the $\ln R$ resummation for {\it inclusive} jet cross sections with a small jet radius $R$. It is important to emphasize again that our approach for {\it inclusive} jet cross sections is different from the usual {\it exclusive} jet production where different types of jet functions enter into the calculations. Finally, we  presented phenomenological applications of such semi-inclusive jet functions for inclusive jet production in $pp$ collisions at the LHC. We matched our $\ln R$ resummation to the fixed NLO results, and produced both NLO+LL$_{R}$ and NLO+NLL$_{R}$ results. We found numerically that NLO+LL$_{R}$ and NLO+NLL$_{R}$ lead to very similar results, and a reduction of $10-20\%$ in the cross section compared with the NLO results for intermediate $R=0.3-0.5$. Our method can be easily generalized to study jet substructure in the case of inclusive jet production~\cite{Kang:2016ehg}.

\acknowledgments
We thank Werner Vogelsang for lots of inspiring discussions, and for providing his NLO jet code for comparison. In addition, we are grateful to P.~Hinderer, C.~Lee, Y.~Q.~Ma, E.~Mereghetti, P.~Pietrulewicz, I.~Scimemi, I.~Stewart, F.~ Tackmann, and W.~Waalewijn for very helpful discussions and useful comments. This work is supported by the U.S. Department of Energy under Contract No.~DE-AC52-06NA25396, and in part by the LDRD program at Los Alamos National Laboratory. 

\bibliographystyle{JHEP}
\bibliography{bibliography}

\end{document}